 \documentclass{emulateapj}
 
\usepackage{graphics,graphicx}
\usepackage{natbib}
\usepackage{epstopdf}
\usepackage{amsmath,amsfonts, amssymb}

\usepackage{amsmath,amsfonts, amssymb}

\usepackage{mathrsfs}

\usepackage{dsfont}

\usepackage{subfigure}
\usepackage{url}
\usepackage{paralist}

\newcommand{\given}{\,|\,}
\newcommand{\transpose}[1]{{#1}^{\mathsf{T}}}
\newcommand{\inverse}[1]{{#1}^{-1}}

\shorttitle{Finding, characterizing and classifying variable sources in multi-epoch sky surveys}
\shortauthors{Hernitschek et al.}

\begin{document}


\title{Finding, characterizing and classifying variable sources in multi-epoch sky surveys:\\
QSOs and RR Lyrae in PS1 3$\pi$ data}


\author{Nina Hernitschek, Edward F. Schlafly, Branimir Sesar and Hans-Walter Rix\altaffilmark{1}}
\email{hernitschek@mpia-hd.mpg.de}

\author{David W. Hogg\altaffilmark{1}\altaffilmark{2}\altaffilmark{3}}   

\author{\v{Z}eljko Ivezi\'{c}\altaffilmark{5}}

\author{Eva K. Grebel\altaffilmark{4}}   

\author{Eric F. Bell \altaffilmark{12}, Nicolas F. Martin \altaffilmark{12,1}, W. S. Burgett\altaffilmark{14}, H. Flewelling\altaffilmark{6}, K. W. Hodapp\altaffilmark{6}, N. Kaiser\altaffilmark{6}, E. A. Magnier\altaffilmark{6}, N. Metcalfe\altaffilmark{8}, R. J. Wainscoat\altaffilmark{6}, C. Waters\altaffilmark{6}
}

\altaffiltext{1}{Max-Planck-Institut f{\"u}r Astronomie, K{\"o}nigstuhl 17, 69117 Heidelberg, Germany}    
\altaffiltext{2}{Center for Cosmology and Particle Physics, Department of Physics, New York University, 4 Washington Place, New York,
NY, 10003, USA}
\altaffiltext{3}{Center for Data Science, New York University, 726 Broadway, 7th Floor, New York, NY, 10003, USA}
\altaffiltext{4}{Astronomisches Rechen-Institut, Zentrum f{\"u}r Astronomie der Universit{\"a}t Heidelberg, M{\"o}nchhofstr. 12-14, 69120 Heidelberg, Germany}
\altaffiltext{5}{University of Washington, Dept. of Astronomy, Box 351580, Seattle, WA 98195}    
  
\altaffiltext{6}{Institute for Astronomy, University of Hawaii at Manoa, Honolulu, HI 96822, USA}

\altaffiltext{7}{Department of Physics, Durham University, South Road, Durham DH1 3LE, UK}

\altaffiltext{8}{Department of Physics and Astronomy, Johns Hopkins University, 3400 North Charles Street, Baltimore, MD 21218, USA}

\altaffiltext{9}{Department of Astrophysical Sciences, Princeton University, Princeton, NJ 08544, USA}

\altaffiltext{10}{US Naval Observatory, Flagstaff Station, Flagstaff, AZ 86001, USA}

\altaffiltext{11}{Department of Physics, Harvard University, Cambridge, MA 02138, USA} 

\altaffiltext{12}{Michigan Institute for Research in Astrophysics, 300E West Hall 1085 S. University Ave. Ann Arbor, MI 48109-1107}
\altaffiltext{13}{Observatoire astronomique de Strasbourg, Universit\'{e} de Strasbourg, CNRS, UMR 7550, 11 rue de l'Universit\'{e}, F-6700 Strasbourg, France}
\altaffiltext{14}{GMTO Corp., 251 S. Lake Ave., Suite 300, Pasadena, CA 91101, USA}



\begin{abstract}In area and depth, the Pan-STARRS1 (PS1) 3$\pi$ survey is 
unique among many-epoch, multi-band surveys and has enormous potential for  
all-sky identification of variable sources. 
PS1 has observed the sky typically seven times in 
each of its five bands ($grizy$) over 3.5 years, but 
unlike SDSS not simultaneously across the bands. 
Here we develop a new approach for quantifying statistical 
properties of non-simultaneous, sparse, multi-color lightcurves through 
light-curve structure functions, effectively turning PS1 into a $\sim 35$-epoch 
survey. We use this approach to estimate variability amplitudes and timescales 
$(\omega_r, \tau)$ for all point-sources brighter than $r_{\mathrm{P1}}=21.5$ mag in
the survey.
With PS1 data on SDSS Stripe 82 as ``ground truth", we use a 
Random Forest Classifier to identify QSOs and RR Lyrae based on their 
variability and their mean PS1 and WISE colors.
We find that, aside from the Galactic plane, QSO and RR Lyrae samples of
purity $\sim$75\% and completeness $\sim$92\% can be selected.
On this basis we have identified a sample 
of $\sim 1,000,000$ QSO candidates, as well as an 
unprecedentedly large and deep sample of $\sim$150,000 RR Lyrae candidates with
distances from $\sim$10 kpc to $\sim$120 kpc. 
Within the Draco dwarf spheroidal, we demonstrate a distance precision of 6\% 
for RR Lyrae candidates.
We provide a catalog of all likely variable point sources and likely QSOs in PS1, a
total of $25.8\times 10^6$ sources.
\end{abstract}


\section{Introduction}
\label{sec:Introduction}

Time domain astronomy is widely held as one of the promising growth areas of astrophysics for the next decade.
Over the last decade, a number of time-domain, wide-area sky surveys with modern digital detectors have been implemented, such as the Palomar Transient Factory Survey \citep[PTF,][]{Rau2009}, Lincoln Near-Earth Asteroid Research \citep[LINEAR,][]{Stokes1998}, the Catalina surveys \citep{Drake2009}, and  Kepler\footnote{\url{http://www.kepler.arc.nasa.gov/}}. In this context, the Pan-STARRS1 survey (PS1) 3$\pi$ \citep{Chambers2011} offers a unique combination of area, time sampling and depth. PS1 data have been extensively used to find and study transient sources, such as supernovae \citep{Rest2014} or episodic black hole accretion \citep{Gezari2012}, focusing mostly on the many-epoch coverage in the medium-deep fields. It lends itself also to finding and characterizing sources of less ephemeral variability,
and can do so across most of the sky. Such sources of interest are, for example, QSOs 
and variable stars, such as RR Lyrae. 

PS1 is a multi-epoch survey that covered three quarters of the sky at typically 35 epochs 
between 2010 and the beginning of 2014. Yet, in any one of its five bands ($g_{\rm P1},r_{\rm P1},i_{\rm P1},z_{\rm P1},y_{\rm P1}$), it is only a few-epoch survey, and the observations in different bands are not taken simultaneously. 

Though there are approaches for finding RR Lyrae in PS1 based on their
variability properties \citep[e.g.][]{Abbas2014a,Abbas2014b}, there are no readily available approaches to exploit the full
information content of the data, e.g. to find, identify, and characterize variable sources generically.

In this paper, we lay out, develop, test, and apply an approach to characterize variable sources in a survey such as PS1. The basic approach should also be very relevant to the Large Synoptic Survey Telescope (LSST)\footnote{LSST Science Collaborations and LSST Project 2009, LSST Science Book, Version 2.0, arXiv:0912.0201, \url{http://www.lsst.org/lsst/scibook}}, which will also collect non-simultaneous multi-band time-domain data. Our methodology encompasses three basic steps: first, identifying sources that clearly vary; second, characterizing their lightcurves with a multi-band structure function; finally, using the identification of variable sources to train an automatic classifier. The last step is carried out using a Random Forest Classifier that takes the classification available for the Sloan Digital Sky Survey (SDSS) Stripe 82 (S82) \citep{Schneider2007,Schmidt2010, Sesar2010} to classify variable sources within PS1 3$\pi$. Throughout this analysis, Stripe 82, which was fully observed by the PS1 survey, serves as a testbed for many aspects of the analysis. 

In the classification analysis, we focus on two classes of astrophysical objects: 
QSOs and RR Lyrae. These objects have numerous applications. For example, the RR Lyrae can act as tracers of the Milky Way's stellar outskirts \citep{Sesar2010, Sesar2013a, Sesar2013b} with high distance precision.
Variability of QSOs is astrophysically interesting for a variety of reasons \citep{Schmidt2010, Morganson2014, Hernitschek2015}, but QSO candidates may
also serve as reference sources for calibrating the astrometry of sources near the Galactic plane. There are many other classes of variables (e.g. Cepheids and other pulsating variables) for which PS1 forms an attractive data base; but we do not attempt an exhaustive variable classification in this paper.

This paper is organized as follows. 
In Section \ref{sec:Data}, we provide a brief description of the PS1 survey, and the time-sampling of its 3$\pi$ sub-survey. We also describe complementary WISE data that prove important for QSO/RR Lyrae discrimination, as well as the existing QSO and RR Lyrae classification in SDSS S82, which is central for training a Random Forest Classifier.
In Section \ref{sec:Methodology}, we describe the methodology that takes us from PS1 lightcurves to QSO and RR Lyrae candidates. We lay out the usage of statistical variability measures, describe the approach of structure functions and state how the classification available for SDSS Stripe 82 helps us in classifying variable objects in PS1 3$\pi$. 
In Section \ref{sec:Results} we demonstrate, relying on Stripe 82 data as ground truth, how well the 
identification and classification of variables with PS1 data works. In particular, we quantify the purity and completeness of various QSO and RR Lyrae samples, as well as discuss results in areas other than Stripe 82, e.g. the Galactic anticenter.
Finally, we provide and describe a catalog of QSO and RR Lyrae candidates across three quarters of the sky. We discuss our results and present conclusions in Section \ref{sec:DiscussionandConclusion}.

\section{Data}
\label{sec:Data}

Our approach for calculating variability measures and using them to detect and classify variable sources is based on PS1 3$\pi$ data, supported by time-averaged photometry from the Wide-field Infrared Survey Explorer (WISE) survey, and sources from SDSS S82 as ground truth. In this section, we describe the pertinent properties of these surveys. From PS1 3$\pi$, we use the derived variability measures as well as mean magnitudes and colors for classification.

\subsection{PS1 3$\pi$ Data}
\label{sec:PS13pidata}

Pan-STARRS is a wide-field optical/near-IR survey telescope system located at Haleakala Observatory on the island of Maui
in Hawaii. The PS1 survey \citep{Kaiser2010} is collecting multi-epoch, multi-color observations undertaking a number of surveys, among which the PS1 3$\pi$ survey \citep{Chambers2011} is the largest. 
It has observed the entire sky north of declination $-30^{\circ}$ in five filter bands ($g_{\rm P1},r_{\rm P1},i_{\rm P1},z_{\rm P1},y_{\rm P1}$) with average wavelengths of 481, 617, 752, 866, and 962 nm, respectively \citep{Stubbs2010,Tonry2012} with a 5$\sigma$ single epoch depth of about 22.0, 22.0, 21.9, 21.0 and 19.8 magnitudes in $g_{\rm P1},r_{\rm P1},i_{\rm P1},z_{\rm P1}, and y_{\rm P1}$, respectively. 
In contrast to the SDSS filters, the $g_{\mathrm{P1}}$ filter extends 20 nm redwards of $g_{\mathrm{SDSS}}$, and the $z_{\mathrm{P1}}$ filter reaches only to 920 nm. PS1 has no $u$ band. In the near-IR, $y_{\mathrm{P1}}$ covers the region from 920 nm to 1030 nm. More detailed descriptions of these filters and their calibration can be found in \citet{Stubbs2010} and \citet{Tonry2012}.

Roughly 56\% of the PS1 telescope observing time was dedicated to the PS1 3$\pi$ survey, with an observing cadence optimized for the detection of near-Earth asteroids and slow-moving solar system bodies. The PS1 3$\pi$ survey plan is to observe each position 4 times per filter per year, where the epochs are typically split into two pairs of exposures per year per band (transit-time-interval (TTI) pairs) taken ${\sim}$25 minutes apart in the same band. This allows for the discovery of moving or variable sources. The sky north of declination $-30^{\circ}$ was planned to be observed 4 times in each band pass per year \citep{Chambers2011}. Through periods of bad weather and telescope downtime in practice, fewer epochs were observed.

Images are automatically processed using the survey pipeline \citep{Magnier2008}, performing bias subtraction, flat
fielding, astrometry, photometry, as well as image stacking and differencing. 
The photometric calibration of the survey is better than one hundredth of a magnitude \citep{Schlafly2012}.

All data processing shown here is carried out under PS1 catalog processing version PV2, where the average number of total detections per source is 55 over 3.7 years, including some data taken in non-photometric conditions.

\subsection{PS1 Object Selection and Outlier Cleaning}
\label{sec:objectselectionandoutliercleaning}

We perform a number of cuts on the PS1 data to remove outliers and unreliable data.  These cuts fall into two categories: \emph{detection} cuts that remove individual detections, and \emph{object} cuts that remove all detections of a source from the analysis.

\subsubsection{Detection Cuts}

The most important detection cut we apply is to remove data taken in non-photometric conditions, according to \citet{Schlafly2012}, and data from any Orthogonal Transfer Array (OTA) where the detections of bright stars on that chip are on average over 0.02 mag too faint.  These cuts remove about 30\% of detections.

The second most important detection cut we apply is to remove observations which land on bad parts of the detector, as indicated by having \texttt{psf\_qf\_perfect} $<$ 0.95.  This removes about 10\% of detections.  Similarly importantly, we exclude any observation where the PSF magnitude is inconsistent with the aperture magnitude by more than 0.1 mag or four times the estimated uncertainties, removing 10\% of detections.

We remove any detections with problematic conditions noted by the PS1 pipeline, according to the detections' flags.   For the cleaning flags used, see Table \ref{tab:cleaningflags} and also \citet{Magnier2013}.  This eliminates only about 2\% of detections. 

Finally, we apply an outlier cleaning based on the $z$-score of the individual measurements $z_i =( m_{i} - \mu(b_i)) / \sigma_{ i}$, where $m_i$ is a given magnitude measurement, $\sigma_i$ is its uncertainty, and $\mu(b_i)$ is the error-weighted mean magnitude of all measurements of that source in its band $b_i$.  This eliminates 2\% of detections, and we limit it to eliminate at most 10\% of the detections of any individual source.

\begin{figure*}[H]
\begin{center}    
\subfigure[]{\includegraphics{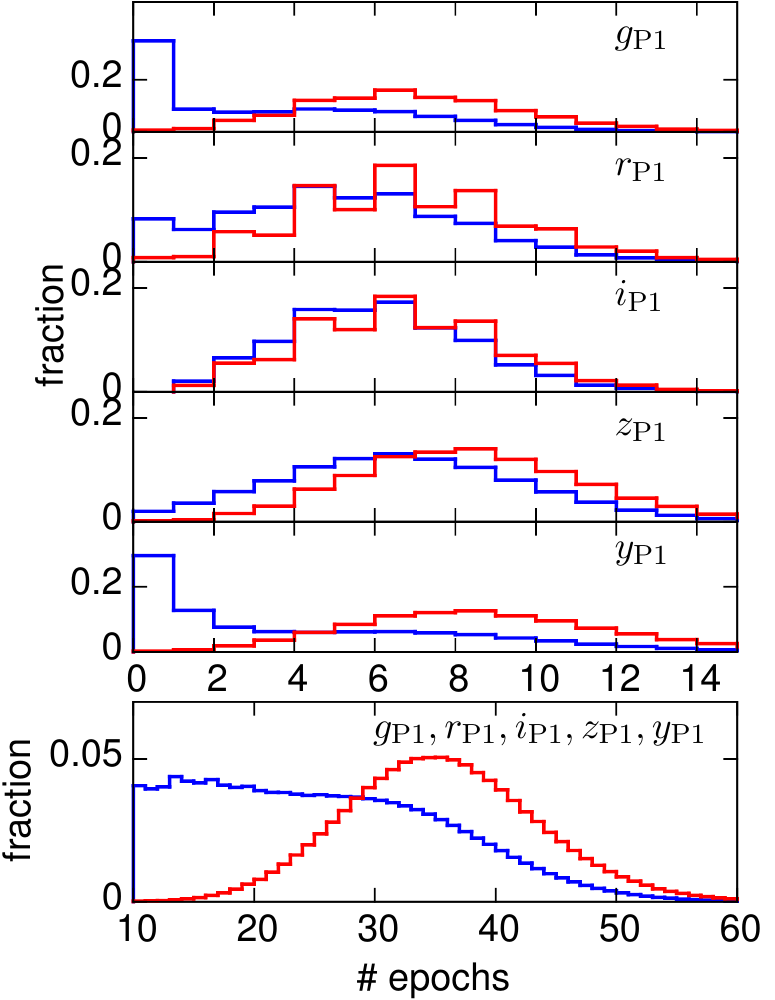}}
\subfigure[]{\includegraphics{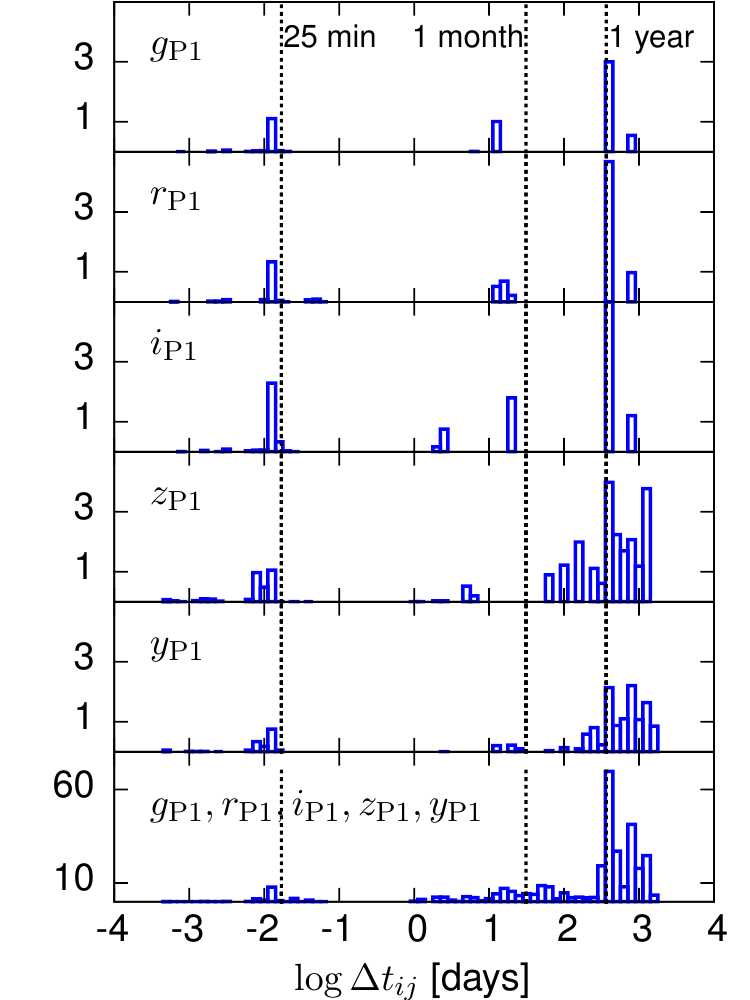}}
\caption{\footnotesize{The typical number of observations (left panels) and the observational cadence (right panels) of PS1 data after source and detection outlier cleaning. a) Average number of epochs in each band for processed objects around the Galactic north pole, after outlier cleaning, for $15<i_{\rm P1}<18$ (red) and $18 \leq i_{\rm P1}<21.5$ (blue). Sources having only few epochs are found only among the faint stars. A minimum number of 10 epochs was enforced by the cleaning. Fractions in the plot are with respect to the total number of sources within $15<i_{\rm P1}<18$ and $18 \leq i_{\rm P1}<21.5$, respectively. b) Average cadence in each band for processed objects for $15<i_{\rm P1}<21.5$ (for 5425 objects around Galactic north pole, after outlier cleaning).}}
\label{fig:PS1dataproperties}
 \end{center}
\end{figure*}

Fig. \ref{fig:PS1dataproperties} gives the number of epochs, as well as their cadence, in each band after all of these cuts have been applied. The average number of surviving epochs per source is 35 rather than the total 55 observations.

The detection cuts we make are summarized in Table~\ref{tab:obsoutliercleaning}. We note that if a detection has one problematic condition, it is more likely than otherwise to also be affected by other problematic conditions.

\subsubsection{Object Cuts}
We also exclude all detections of some objects from consideration.  To ensure that we consider only objects with enough epochs and high enough signal to noise to be appropriate for variability studies, we select only objects having
\begin{compactenum}[(i)]
\item $15 < \langle g_{\rm P1} \rangle, \langle r_{\rm P1} \rangle, \langle i_{\rm P1} \rangle < 21.5$, where $\langle \cdot \rangle$ is the error-weighted mean magnitude after applying detection cuts
\item at least 10 epochs remaining after after applying detection cuts
\end{compactenum}
We have imposed two additional criteria to remove extended objects, as well as
objects thought to have problematic PS1 detections:
\begin{compactenum}[(i)]
\setcounter{enumi}{2}
\item fewer than 25\% of epochs eliminated by \texttt{psf\_qf\_perfect}$\leq$0.95
\item fewer than 25\% of epochs eliminated by $|$\texttt{ap\_mag} - \texttt{psf\_inst\_mag}$| \geq$ max($4\sigma$, 0.1).
\end{compactenum}

Among sources within a magnitude range of 15 to 21.5, these two criteria each remove about $5\%$ of objects.  This was significantly more than expected.  However, visual inspection of a selection of affected sources indicates that these cuts were unnecessarily restrictive. These sources could have in fact been included in the analysis without difficulty, but for now we accept the loss. We term this loss a ``selection loss'', and note that it means that our catalogs (QSOs, RR Lyrae, and variable objects in general) will be missing 10\% of all objects.

More than $3.88 \times 10^ 8$ objects across three quarters of the sky survive these cuts, and we analyze the variability of all of them.

\def\arraystretch{1.5}

\begin{deluxetable*}{p{6cm}p{1.7cm}p{8cm}}
\tabletypesize{\scriptsize}
\tablecaption{Bit-flags used to exclude bad or low-quality detections \label{tab:cleaningflags}}
\tablewidth{0pt}
\tablehead{
\colhead{FLAG NAME} & \colhead{Hex Value} & \colhead{Description}}
\startdata
PM\_SOURCE\_MODE\_FAIL & 0x00000008  & Fit (non-linear) failed (non-converge, off-edge, run to zero) \\
PM\_SOURCE\_MODE\_POOR & 0x00000010 &  Fit succeeds, but low-SN or high-Chisq \\
PM\_SOURCE\_MODE\_SATSTAR & 0x00000080  & Source model peak is above saturation \\
PM\_SOURCE\_MODE\_BLEND & 0x00000100  & Source is a blend with other sources \\
PM\_SOURCE\_MODE\_BADPSF & 0x00000400 &  Failed to get good estimate of object's PSF \\
PM\_SOURCE\_MODE\_DEFECT & 0x00000800 &  Source is thought to be a defect \\
PM\_SOURCE\_MODE\_SATURATED & 0x00001000 &  Source is thought to be saturated pixels (bleed trail) \\
PM\_SOURCE\_MODE\_CR\_LIMIT & 0x00002000 &  Source has \texttt{crNsigma} above limit \\
PM\_SOURCE\_MODE\_MOMENTS\_FAILURE & 0x00008000  &  could not measure the moments \\
PM\_SOURCE\_MODE\_SKY\_FAILURE & 0x00010000  & could not measure the local sky \\
PM\_SOURCE\_MODE\_SKYVAR\_FAILURE & 0x00020000 & could not measure the local sky variance \\
PM\_SOURCE\_MODE\_BIG\_RADIUS   & 	 0x00100000 & poor moments for small radius, try large radius \\
PM\_SOURCE\_MODE\_SIZE\_SKIPPED & 0x10000000 &  size could not be determined \\
PM\_SOURCE\_MODE\_ON\_SPIKE & 0x20000000 &  peak lands on diffraction spike \\
PM\_SOURCE\_MODE\_ON\_GHOST & 0x40000000 &  peak lands on ghost or glint \\
PM\_SOURCE\_MODE\_OFF\_CHIP & 0x80000000 &  peak lands off edge of chip
\enddata
\end{deluxetable*}

\begin{deluxetable*}{p{5.5cm}c}
\tabletypesize{\scriptsize}
\tablecaption{Cuts used to exclude bad detections \label{tab:obsoutliercleaning}}
\tablewidth{0pt}
\tablehead{
\colhead{Condition} & \colhead{Fraction of detections removed} }
\startdata
Photometric conditions & 0.29\\
$|$\texttt{ap\_mag} - \texttt{psf\_inst\_mag}$|<$ max(4$\times \sigma_m$, 0.1) & 0.10\\
\texttt{psf\_qf\_perfect} $>$ 0.95 & 0.11\\
Pipeline flags (Tab. \ref{tab:cleaningflags}) & 0.017   \\    
$\left| z_i - z_{\mathrm{median}} \right| <5\sigma$ & 0.02\\
\enddata
\end{deluxetable*}


\subsection{WISE Data}
\label{sec:WISEdata}

WISE (Wide-field Infrared Survey Explorer) is a NASA infrared-wavelength astronomical space telescope providing mid-infrared data with far greater sensitivity than any previous survey. It performed an all-sky survey with imaging in four photometric bands over ten months \citep{Wright2010}. \citet{Nikutta2014} have shown that the color $W12=W1-W2>0.5$ is an excellent criterion to isolate QSOs, because $W12$ is an indicator of the hot dust torus in AGN. To aid in the QSO identification, we want to find objects with these unusual W12 colors, but need to make sure that these colors are not merely a consequence of poor WISE photometry.
For objects with good measurements ($\sigma_{W1}<0.3$, $\sigma_{W2}<0.3$), we use $W12$ as parameter for classification. $1.46 \times 10^8$ of the $3.88 \times 10^ 8$ selected objects from Sec. \ref{sec:PS13pidata} have reliable $W12$ ($\sigma_{W1}<0.3$, $\sigma_{W2}<0.3$, where $\sigma_{W1}$, $\sigma_{W2}$ are the errors given on the WISE magnitudes).

\subsection{SDSS S82 Sources}
\label{sec:SDSSdata}

The Sloan Digital Sky Survey \citep[SDSS,][]{York2000} is a major multi-filter imaging and spectroscopic survey using a dedicated 2.5-m wide-angle optical telescope at Apache Point Observatory in New Mexico, United States. 
The Sloan Legacy Survey covers about 7,500 degrees of the Northern Galactic Cap in optical $ugriz$ filters with average wavelengths of 355.1, 468.6, 616.5, 748.1 and 893.1 nm. In typical seeing, it has a 95\% completeness down to magnitudes of 22.0, 22.2, 22.2, 21.3, and 20.5, for $u$, $g$, $r$, $i$, $z$, respectively.
Additionally, the Sloan Legacy Survey contains three stripes in the South Galactic Cap totaling 740 square degrees. The central stripe in the South Galactic Cap, Stripe 82 (S82), was scanned multiple times to enable a deep co-addition of the data and to enable discovery of variable objects.
S82 has ${\sim}$60 epochs of imaging data in $ugriz$, taken over ${\sim}$5 years, where extensive spectroscopy provides a reference sample of nearly 10,000 spectroscopically confirmed quasars \citep{Schneider2007,Schmidt2010}.
For S82, there is also a sample of 483 identified RR Lyrae available \citep{Sesar2010}.

The classification of QSOs and RR Lyrae in SDSS S82 will be used as a ground truth. This means, they will be used as training set for classification as well as for testing how well our classification method works (see Section \ref{sec:Methodology}).

Within $-50^{\circ} < \alpha < 60^{\circ}$, $-1.25^{\circ} < \delta < 1.25^{\circ}$, there are 9073 QSO and 482 RR Lyrae from the samples mentioned above. Out of these, 7633 QSO and 415 RR Lyrae are cross-matched to objects in our PS1 selection. We also select more than $1.85 \times 10^ 6$ "other" objects from S82.
10\% of these are missing because of the cuts of Section \ref{sec:objectselectionandoutliercleaning}, and the remaining objects are outside our magnitude range of interest.

\section{Methodology}
\label{sec:Methodology}

\begin{figure*}
              \includegraphics[width=1.0\textwidth]{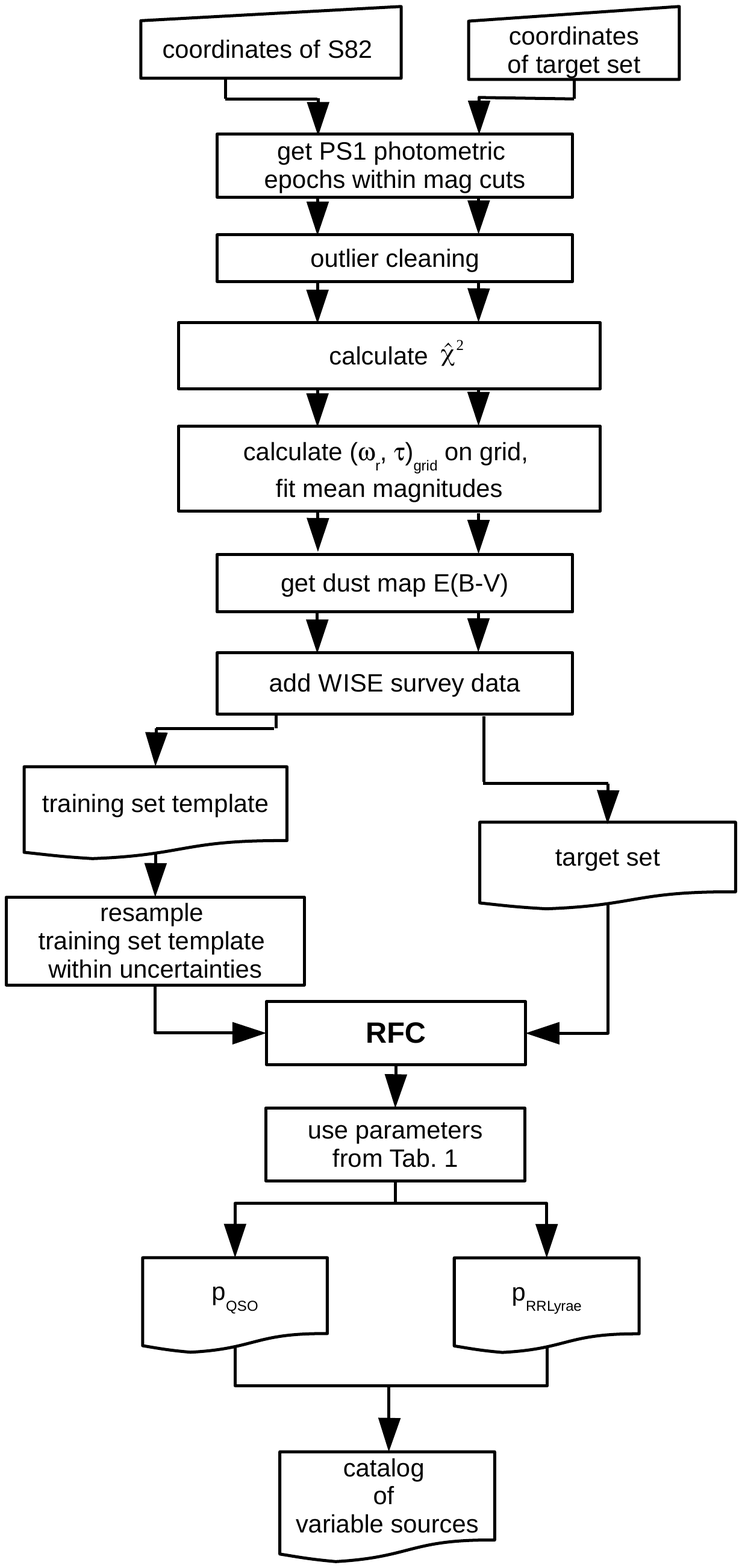}    
                \caption[shortcaption]{\footnotesize{Logic flowchart for finding and classifying variable sources as set out in Section \ref{sec:Methodology}}.}
                \label{fig:flowchart}
\end{figure*}

In this section we describe the three steps we take to identify and characterize variable point sources: first, determine whether sources are 
variable; second, characterize their variability with a structure function; and third, attribute classifications. Classification is carried out using a Random Forest Classifier that utilizes a training set from SDSS S82. 
Throughout the following steps we assume that all data conform to the selection requirements described in Section \ref{sec:Data}.
Fig. \ref{fig:flowchart} illustrates the logical flow of the methodology that is detailed in the following subsections.

\subsection{Identifying Significantly Varying Sources}
\label{sec:varibilityq}
\begin{figure*}
\begin{center}    
\subfigure[]{\includegraphics{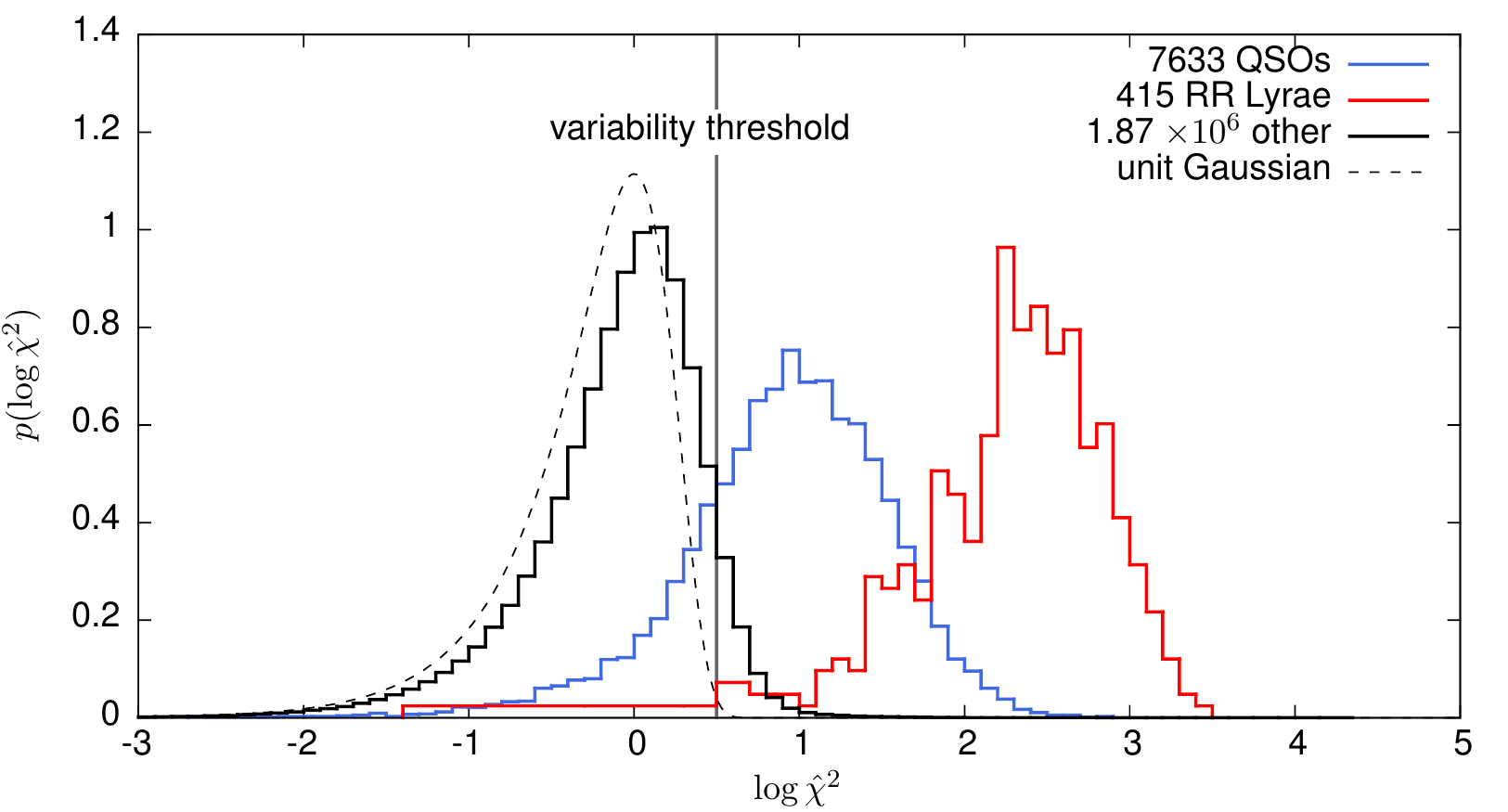}}
\subfigure[]{\includegraphics{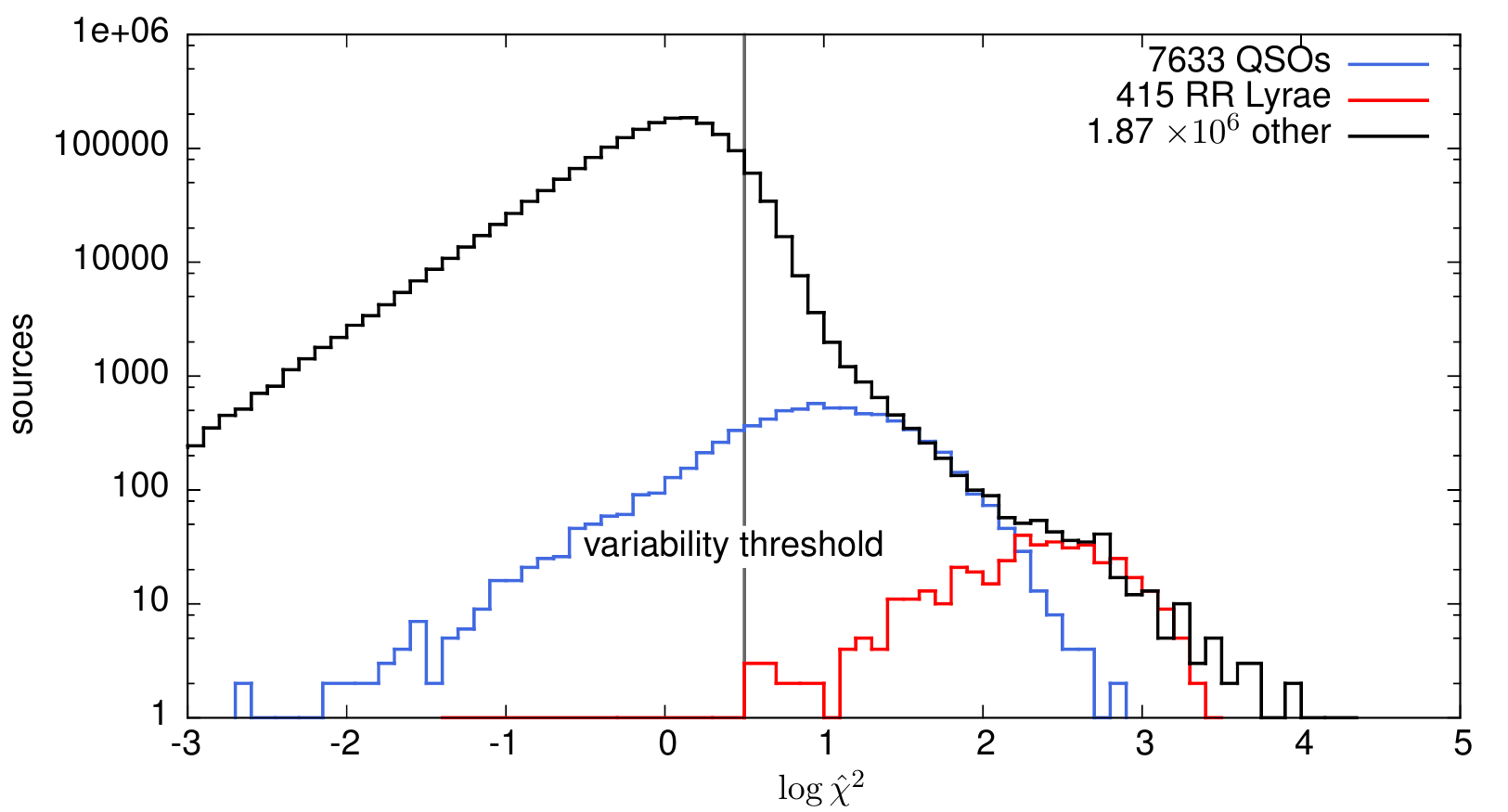}}
\caption{\footnotesize{Histograms for $\hat{\chi}^2$ of the training set's sources after outlier cleaning; PS1 photometry in S82 region, type from SDSS.\newline
(a) Normalized histogram, overplotted: theoretical expectation from unit Gaussian distribution $(\mu=0,\sigma=1)$. The differences between the black histogram and the dashed line arises from a combination of noise-model imperfections and actual variability of objects. The cutoff for the variability criterion ($\log \hat{\chi}^2>0.5$, see Sec. \ref{sec:Catalog}) is given as a grey line.\newline
(b) Full histogram showing how the distribution of ``other" sources superimposes the distribution of QSOs and RR Lyrae due to the high number of ``other" sources.}}
\label{fig:chihatdistribution}
\end{center}
\end{figure*}

We start by laying out a very generic and non-parameteric measure for variability, simply to characterize the significance of
variability by a scalar quantity. Specifically, we define
 \begin{equation}
\hat{\chi}^2=\frac{\chi^ 2_{\mathrm{source}} -N_{d.o.f}}{\sqrt{2 \, N_{d.o.f}}},
\label{eqn:q}
\end{equation}
with
\begin{equation}
\chi^ 2_{\mathrm{source}}  = \sum_{\lambda} \sum^N_{i=1} \frac{(m_{\lambda,i} - \langle m_{\lambda}\rangle)^ 2}{{\sigma}_{\mathrm{\lambda},i}^2}
\end{equation}
where $N$ is the total number of photometric points for one object across all $n$ bands, 
the sum over $\lambda$ is over the PS1 bands $g_{\rm P1}$,$r_{\rm P1}$,$i_{\rm P1}$,$z_{\rm P1}$,$y_{\rm P1}$, and $N_{d.o.f}=N-n$ is the number of degrees of freedom.\newline
Assuming that most of the sources are not variable, we expect the distribution of $\hat{\chi}^2$ to be a unit Gaussian distribution. In contrast, varying sources should form a ``tail'' of higher $\hat{\chi}^2$.  Figure \ref{fig:chihatdistribution} shows the normalized distribution of $\hat{\chi}^2$, derived from the PS1 photometry of all selected objects in S82, with known QSOs (blue) and known RR Lyrae (red) shown
in separate (normalized) distributions.
The ``other'' objects have a $\hat{\chi}^2$-distribution close to that expected for non-varying sources (dashed line), confirming that most sources in the sky are non-varying (within a level of less than a few percent) and that the PS1 photometry is reliable. 
The QSOs and RR Lyrae appear well separated in the {\it normalized} distibutions. However, there are only 415 RR Lyrae and 7630 QSO, compared to ${\sim}1.85 \times 10^6$ ``other'' objects in SDSS S82 cross-matched to PS1 and surviving the cuts of Sec. \ref{sec:objectselectionandoutliercleaning}. Fig. \ref{fig:chihatdistribution} (b) shows how the distribution of ``other" sources superimposes the distribution of QSOs and RR Lyrae due to the high number of ``other" sources. 

Therefore, a simple criterion such as $\hat{\chi}^2$ is insufficient to identify QSOs or RR Lyrae.
In the subsequent analysis, all objects are used, though for RR Lyrae only, one could in principle restrict oneself to objects with $\hat{\chi}^2>10$ without losing completeness.

\subsection{Non-simultaneous, Multi-band Structure Functions}
\label{subsec:multi-band-structure-function}

Beyond simply establishing variability, variable sources can and
should be characterized by the amplitude of their variability and the timescales over which they vary. 
A useful and well-established tool for this is the structure function \citep{Hughes1992, Collier2001, Kozlowski2010}: it gives
the mean squared magnitude difference (in a given band) between pairs of observations of some object's brightness ($\Delta m$) as a function of the time lag between
the observations ($\Delta t$). There are various ways to parameterize such a structure function,
and the damped random walk (DRW) is a useful function family. For a DRW, the structure function 
is specified by two parameters, $\tau$ and $V_{\infty}$, and is given by

\begin{align}
\begin{split}
V(\Delta t| \tau,V_{\infty}) = V_{\infty}(1-e^{-|\Delta t|/\tau})  \, .
\label{eq:strucfunc}
\end{split}
\end{align}

In this notation $V(\Delta t)$ reflects the expectation value for the squared 
magnitude difference, $\Delta m^2$, among measurements separated in time by $\Delta t$;  $V(\Delta t)$ is simply $\sqrt{2}$ times the expected magnitude variance during $\Delta t$. 
 $V_{\infty}$ is conventionally denoted as $\omega^ 2$, and 
$\tau$ is called the decorrelation time of the DRW.  The source variability is then characterized by two structure function parameters, $\omega$ and $\tau$.

Objects of different classes typically occupy different regions in structure function parameter space. As we show below, the likelihood  $p(\vec{m} \given \mathrm{SF\ parameters})$ can be used to select remarkably pure and complete samples of QSOs as well as RR Lyrae, which makes selection by structure function parameters an efficient approach for both selecting stochastically varying and periodic variable objects \citep{Schmidt2010}.

The cadence of surveys like the SDSS provides data that allows application of the usual single-band formulation of structure functions. However, the cadence of PS1 3$\pi$ data, which observes in different bands at different epochs (see Sec. \ref{sec:Data}), makes it necessary to extend this approach for multi-band fitting. We need a practical approach to turn the ${\sim} 6-9$ epoch PS1 light curves in each band into a ${\sim} 35$ epoch overall light curve.
If objects were to vary the same way in all observed bands, implementing such an approach would simply amount to determining the (time-averaged) mean color of the object and shifting the light curves in the different bands to a common magnitude. However, in practice, most astrophysical objects vary more at shorter wavelengths. To account for this, the multi-band model we present here has, beyond $\omega$ and $\tau$, a set of temporal mean magnitude parameters in each PS1 band, $\vec{\mu}$, and it links the variability \textit{amplitudes} $\omega(b)$ in different bands $b$ by a power law with exponent $\alpha$.  Specifically,
\begin{align}
\alpha = \frac{\log(\omega(b)/\omega(r))}{\log(\lambda_b/\lambda_r)} \, ,
\label{eq:alpha}
\end{align}
where $\lambda_b$ is the effective wavelength of the band $b$.

To assign a likelihood to an object's photometry, given a structure function model, we make
use of a Gaussian Process formulation for stochastic source variability. 
In contrast to single-band structure function models \citep[e.g.][]{Rybicki1992,Zu2011,Hernitschek2015}, the Gaussian Process is not applied to any particular band
but instead to an arbitrarily constructed fiducial band which can be scaled and shifted
onto the particular bands. This permits simultaneous treatment of
multiple bands, even when the bands are not observed
simultaneously at the same epochs. It is key in this context to realize that the fiducial band is a latent
variable -- it is never directly observed; only the scaled and shifted
versions are observed, where substantial measurement noise is present.

The fiducial light curve can be described with a zero-mean and unit characteristic variance Gaussian Process.
That is, the prior probability distribution function (pdf) for a set of $N$ fiducial ``magnitudes'' $\vec{q}$ that are
instantiated at observed times $t_n$ is a multivariate normal distribution:
\begin{equation}
p(\vec{q}) = \mathcal{N}(\vec{q}\given 0,C^q) \, ,
\end{equation}
where $C^q$ is a $N\times N$ symmetric positive definite covariance matrix.  In the case of a DRW model, $C^q$ is given by
\begin{equation}
C^{q}_{nn'} = \exp\left[-\frac{|t_n - t_{n'}|}{\tau}\right] \, .
\label{eq:singlebandcovariancematrix}
\end{equation}

This is identical to the usual single band DRW covariance matrix,
except that we have dropped a scale factor $\omega^2$ from Equ. \eqref{eq:singlebandcovariancematrix},
because we have defined the fiducial band $q$ to have unit variance.
This factor reappears in our multi-band structure function through the
scale factors that link the fiducial band to observed bands.

We consider now a given source for which we have $N$ observations across $N_\mathrm{band}$ different bands. The data consist of the magnitude and uncertainty vectors $\vec{m}$ and $\vec{\sigma}$, the times of observation $t_n$, and the corresponding bands $b_n$.  The source also has $N_\mathrm{band}$ temporal mean magnitudes $\vec{\mu}$.  We define the $N \times N_\mathrm{band}$ matrix $\mathbb{M}$ so that
\begin{equation}
\mathbb{M}\vec{\mu} = [\mu(b_1), \mu(b_2), \cdots , \mu(b_N)] \, .
\end{equation}

The likelihood of an individual measurement $m_n$, given its observational uncertainty $\sigma_n$ and a
value for the corresponding fiducial magnitude $q_n$, is found by
shifting and scaling the fiducial magnitude and adding Gaussian noise.
This makes the single-datum likelihood
\begin{equation}
p(m_n\given q_n,b_n,\sigma_n^2) = \mathcal{N}(m_n\given \omega(b_n)q_n + \mu(b_n) , \sigma_n^2),
\end{equation}
where $\omega(b_n)$ is the variability in bandpass $b_n$ relative to the unit variability of the unobserved fiducial band.

Introducing the diagonal $N \times N$ matrix $\Omega$, defined by $\Omega_{ii} = \omega(b_i)$,
the full likelihood is given by
\begin{equation}
p(\vec{m} \given \vec{q},\Sigma) = \mathcal{N}(\vec{m} \given \Omega\vec{q} + \mathbb{M}\vec{\mu}, \Sigma^2) \, ,
\end{equation} 
where $\Sigma$ is a diagonal matrix with $\Sigma_{ii} = \sigma_i$.
Because everything is Gaussian, the latent
fiducial magnitudes never have to be explicitly inferred;
they can all be marginalized out analytically. This marginalization leads to the likelihood given the model, and the covariance matrix of the data:
\begin{eqnarray}
p(\vec{m} \given \mathrm{SF\ parameters}, \vec{\mu}) &= \mathcal{N}(\vec{m}\given \mathbb{M}\vec{\mu} , C) 
\label{eq:likestart} \\
C &= \Omega C^q \Omega + \Sigma^2 \, .
\label{eq:likeend}
\end{eqnarray}
This is identical to the case of a single-band DRW model, except the rows and columns of $C^q$ are scaled by amplitudes $\omega(b_{n})$, $\omega(b_{n'})$ for the bands $b_n$ and $b_{n'}$, and a contribution from the photometric uncertainties is added to the diagonal:
\begin{equation}
C_{nn'}  = \omega(b_n)\,\omega(b_{n'})\,\exp \left[-\frac{|t_n - t_{n'}|}{\tau} \right] + \sigma_n^2\,\delta_{nn'}.
\label{eq:multibandc}
\end{equation}

Equations \eqref{eq:likestart} through \eqref{eq:multibandc}
provide a method for computing the probability of any set of observed magnitudes, given their meta data and the parameters $\omega(b)$, $\tau$, and $\vec{\mu}$.

We are primarily interested in the structure function parameters and are relatively uninterested in the mean magnitudes $\vec{\mu}$.  This is exactly the same situation as in \cite{Zu2011}.  Following that work, the likelihood of the structure function parameters, given the data, marginalized over $\vec{\mu}$, is given by:
\begin{equation}
\label{eq:marginallikelihood}
p(\vec{m} \given \mathrm{SF\ parameters}) = \mathcal{L} \propto |C|^{-1/2}|C_{\mu}|^{1/2}\exp\left(-\chi^2/2 \right) \,
\end{equation}
where
\begin{align}
\begin{split}
C_{\mu} & = (\mathbb{M}^\mathrm{T} C^{-1}\mathbb{M})^{-1} \, \\
\chi^2 & = \transpose{(\vec{m} - \mathbb{M}\vec{\mu})}C^{-1}(\vec{m}-\mathbb{M}\vec{\mu}) \, .
\end{split}
\end{align}
We note that the factor of $|C_\mu|^{1/2}$ in Equation~\eqref{eq:marginallikelihood} comes from the marginalization over $\vec{\mu}$.  We maximize $p(\vec{m} \given \mathrm{SF\ parameters})$ to obtain best fit values of the structure function parameters.  We then obtain $\vec{\mu}$ as the maximum likelihood values of $\vec{\mu}$ given the structure function parameters.  That is, the mean magnitudes are given by
$$ \vec{\mu} = (\mathbb{M}^T C^{-1} \mathbb{M})^{-1} \mathbb{M}^\mathrm{T}C^{-1} \vec{m} \, ,  $$
and have variance $C_\mu$.

\subsection{Interpolating Multi-Band Light Curves with Uncertainties}

One advantage of this approach is that it can be used to predict unobserved data based on observed data.  Because both the process is Gaussian and the
noise is assumed to be Gaussian, conditional predictions of the magnitudes can be made given the observed data and the structure function.  The analysis is exactly the same as in \cite{Rybicki1992}, with the exception that we adopt the multi-band structure function $C$ of Equation~\eqref{eq:multibandc}.  The magnitudes $\tilde{m}_k$ at $K$ unmeasured times $t_k$, taken through bandpasses $b_k$, conditioned on the data in hand, are given by:

\begin{eqnarray}
p(\tilde{m}|\vec{m}) &=& \mathcal{N}(\tilde{m}|\tilde{\mu},\tilde{C})\\
\tilde{\mu} &=& \vec{\nu} + X\cdot\inverse{C}\cdot [\vec{m} - \mathbb{M}\vec{\mu}] \label{eqn:strucfuncfitted}\\
\tilde{C} &=& Y - X\cdot\inverse{C}\cdot\transpose{X}\label{eqn:strucfuncvariance}\\
\vec{\nu} &=& [\mu(b_1), \mu(b_2), \cdots , \mu(b_K)].
\end{eqnarray}

In the case of a multi-band DRW model,
\begin{eqnarray}
X_{kn} &=& \omega(b_k)\,\omega(b_n)\,\exp \left[ -\frac{|t_k - t_n|}{\tau}\right]
\\
Y_{kk'} &=& \omega(b_k)\,\omega(b_{k'})\,\exp \left[-\frac{|t_k - t_{k'}|}{\tau}\right].
\label{eqn:ykk}
\end{eqnarray}

Here $\tilde{m}$ is the column vector of conditional predictions,
$\tilde{\mu}$ and $\tilde{C}$ are a conditional mean vector and a
conditional variance matrix, (temporary) mean vector $\nu$ is
$K$-dimensional, and the matrices $\tilde{C}$, $X$, and $Y$ are $N\times N$,
$K\times N$, and $K\times K$ respectively.  Vectors $\vec{m}$ and $\vec{\mu}$ and
matrix $C$ are defined in Section \ref{subsec:multi-band-structure-function}.

\subsection{Application to PS1 Data}

We have described a general technique for determining structure functions for multi-band, non-simultaneous data.  The key ingredient is a description of the ratios of the variabilities in the different bands, which we characterize by a power law with exponent $\alpha$ (Equation \eqref{eq:alpha}).  The other elements of the structure function analysis -- overall variability, time scale, and linear nuisance parameters (mean magnitudes) -- are the same as in the single band case.  For the case of the PS1 data at hand, it turns out that $\alpha$ is poorly constrained for individual objects, making it preferable
to derive an external estimate of $\alpha$ from other data (SDSS S82), and fix it for the subsequent PS1 analysis.
Assuming Equ. \eqref{eq:alpha}, we used data from SDSS S82 to derive characteristic values for $\alpha$. We found $\alpha \approx -0.65$ for QSOs and $\alpha \approx -1.3$ for RR Lyrae, both with an uncertainty of 0.01, which is in good agreement with \citet{Sesar2012}.  We experimented fitting PS1 data with both choices of $\alpha$, and obtained similarly good fits.  Accordingly, we decided to adopt a single fixed $\alpha = -0.65$ throughout this analysis.  This choice of $\alpha$ corresponds to variability ratios $\omega(b) / \omega(r) = 1.175, 1.00, 0.88, 0.80, 0.75$, where $b$ represents the PS1 bands $g_{\rm P1}$, $r_{\rm P1}$, $i_{\rm P1}$, $z_{\rm P1}$, and $y_{\rm P1}$.

With $\alpha$ fixed, our fit to each source is described by the time scale $\tau$, an overall variability scaled to the $r$ band, $\omega_r$, and the mean magnitudes $\vec{\mu}$.  We calculate these from the {\it PS1 photometry} of all sources within the SDSS S82 area.

\begin{figure*}
\begin{center}    
\subfigure{\includegraphics[trim=2inch 0.2inch 0.0inch 0.0inch, clip=true]{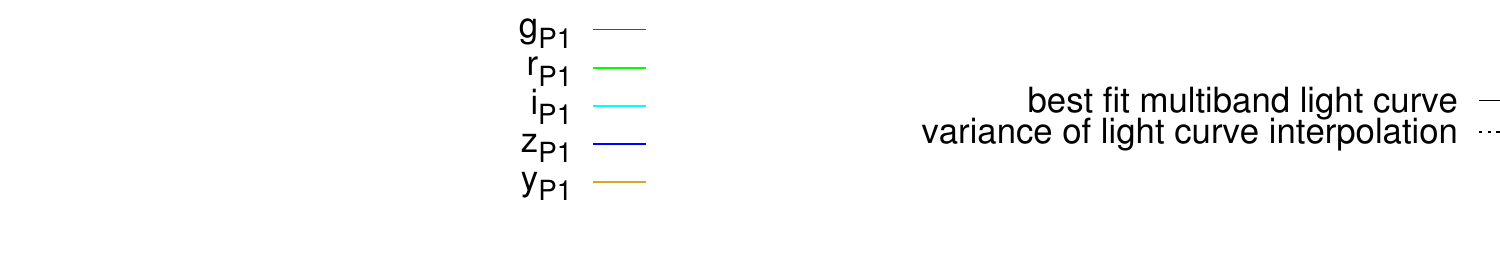}}
    \setcounter{subfigure}{0}
\subfigure[QSO, $\omega_r$=0.13 mag, $\tau$=593 days]{\includegraphics[trim=0.0inch 0.03inch 0.0inch 0.05inch, clip=true]{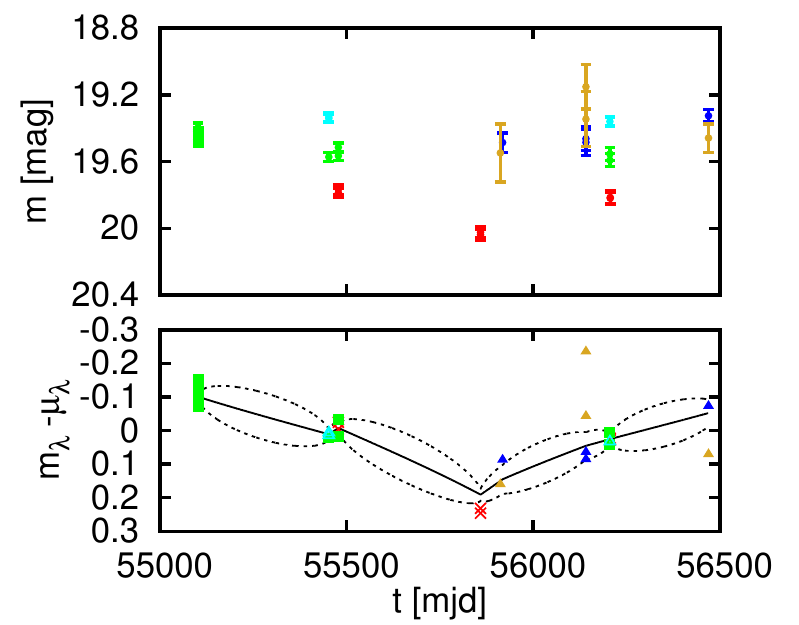}}
\subfigure[RR Lyrae, $\omega_r$=0.2 mag, $\tau$=1.0 days]{\includegraphics[trim=0.0inch 0.03inch 0.0inch 0.05inch, clip=true]{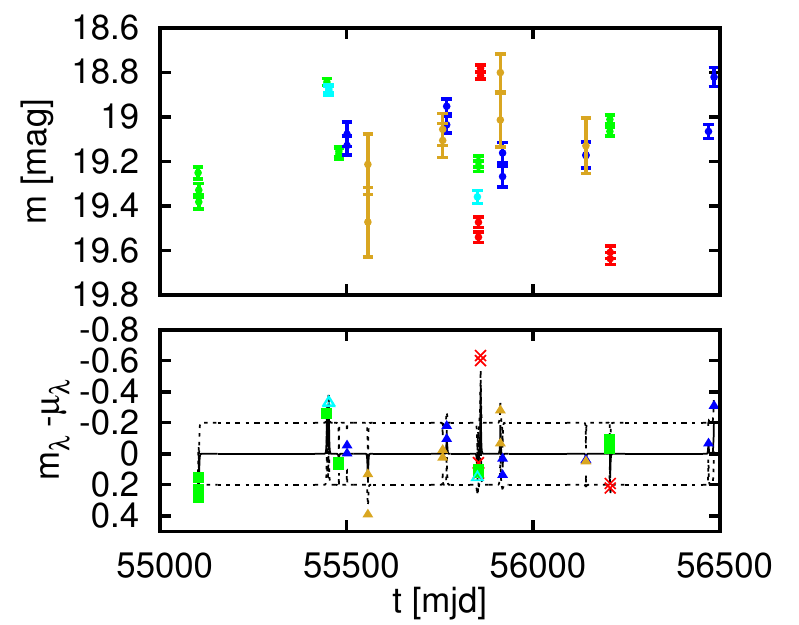}}
\subfigure[other (variable?), $\omega_r$=0.05 mag, $\tau$=496 days]{\includegraphics[trim=0.0inch 0.03inch 0.0inch 0.05inch, clip=true]{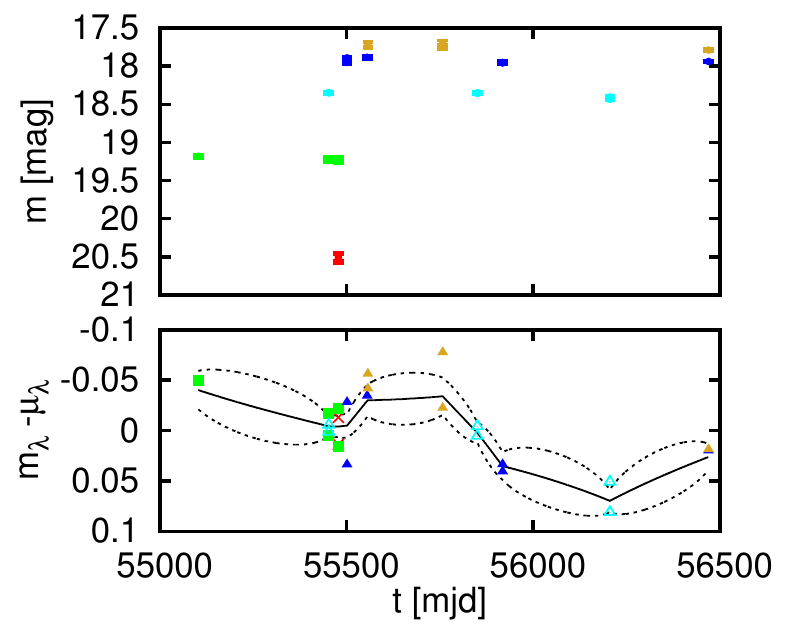}}  
\subfigure[other (nonvariable?), $\omega_r$=0.04 mag, $\tau$=4.3 days]{\includegraphics[trim=0.0inch 0.03inch 0.0inch 0.05inch, clip=true]{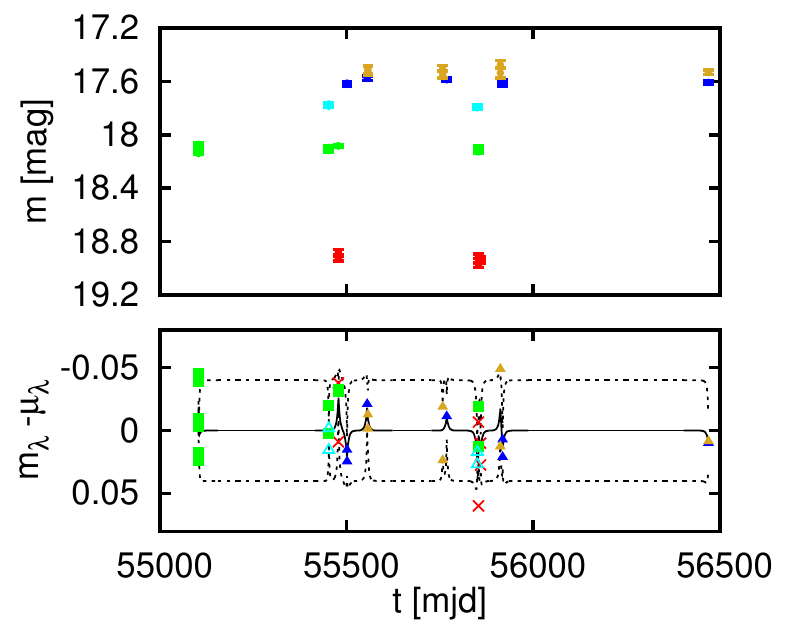}}
\caption{\footnotesize{Examples of multi-band lightcurve models for different types of sources. In each figure, the upper panel gives the PS1 lightcurve data points with error bars after outlier cleaning. The lower panel shows the lightcurve fit by a multi-band DRW structure function. The solid lines represent the best fit mean model lightcurve Equ. \eqref{eqn:strucfuncfitted}. The area between the dotted lines represents the variance Equ. \eqref{eqn:strucfuncvariance} for the $r$ band. For $\omega_r$ and $\tau$,
we use the best MCMC point-estimates of the parameters for each source.
 }}
\label{fig:lcfits}
 \end{center}
\end{figure*}

Fig. \ref{fig:lcfits} shows for example fits to the PS1 photometry of objects in SDSS S82: one QSO, one RR Lyrae, one ``other'' variable object and one seemingly non-varying object. For each object we show the light curve as observed in the five bands (top panel), and the combined light curve after shifting each band by the estimated $\mu(b)$; the structure function parameters $\omega_r$ and $\tau$ are listed for each case. Note that the QSO in Fig. \ref{fig:lcfits} (a) has $\tau$ of over a year, while the RR Lyrae in panel (b) has a $\tau$ of about a day. The Figure also shows the interpolated light curves, given the observations and the structure function parameters, according to the technique of \cite{Rybicki1992}.

\begin{figure*}[H]
\begin{center}    
\subfigure[QSO, $\omega_{r,\mathrm{grid}}$=0.15 mag, $\tau_{\mathrm{grid}}$=660 days]{\includegraphics[]{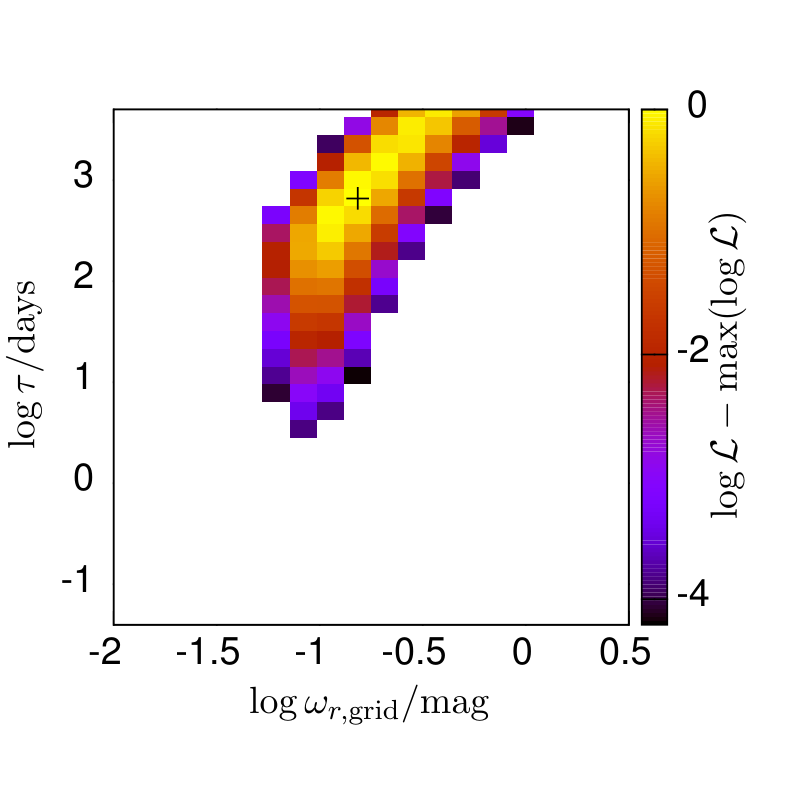}}
\subfigure[RR Lyrae, $\omega_{r,\mathrm{grid}}$=0.28 mag, $\tau_{\mathrm{grid}}$=3.4 days]{\includegraphics[]{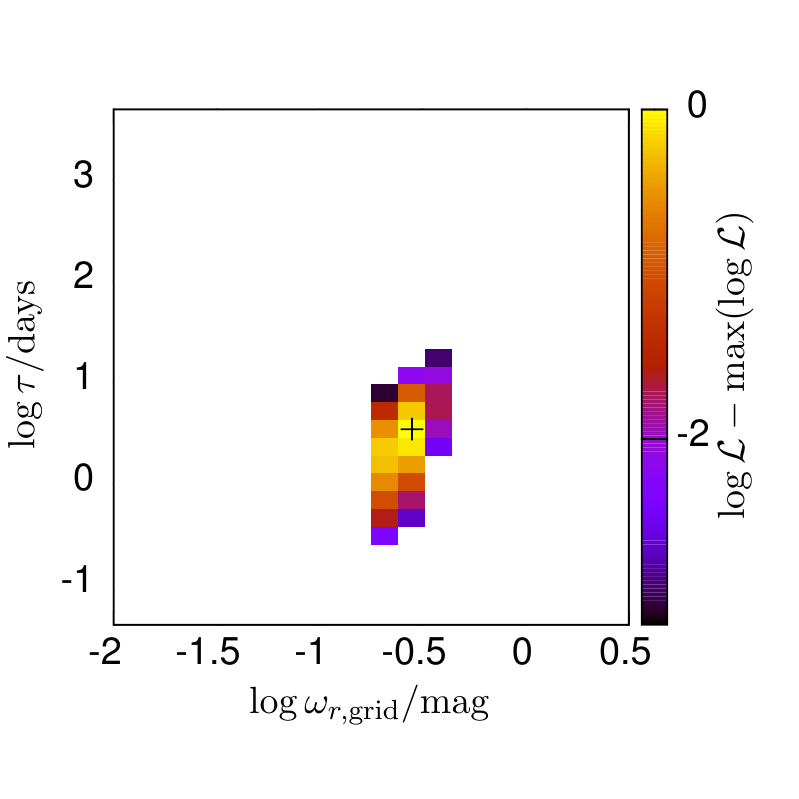}}
\subfigure[other (variable?), $\omega_{r,\mathrm{grid}}$=0.06 mag, $\tau_{\mathrm{grid}}$=990 days]{\includegraphics[]{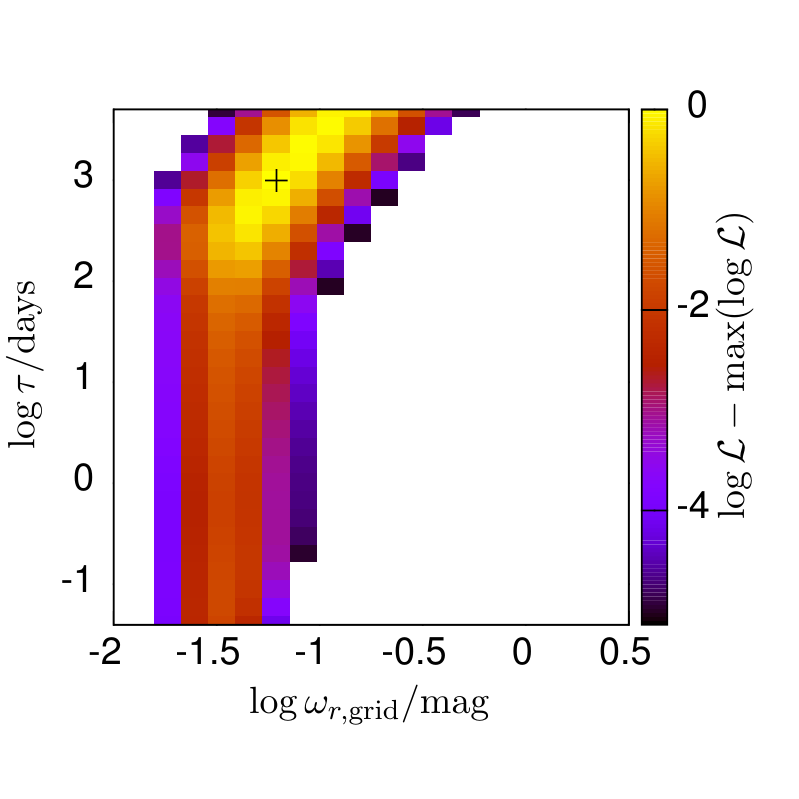}}
\subfigure[other (nonvariable?), $\omega_{r,\mathrm{grid}}$=0.02 mag, $\tau_{\mathrm{grid}}$=0.04 days]{\includegraphics[]{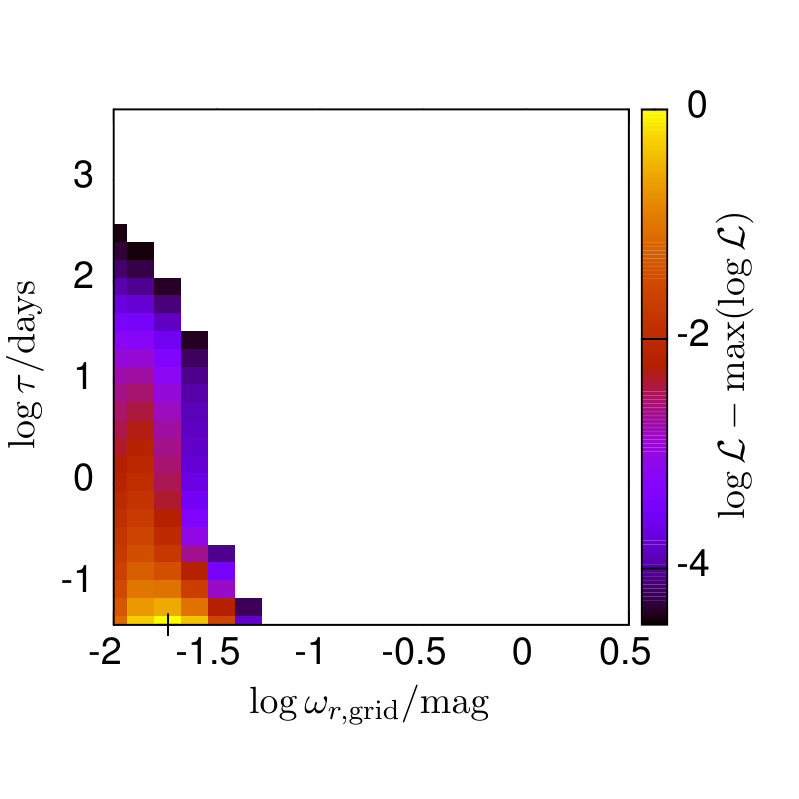}}
\caption{\footnotesize{Gridded log-likelihood estimates for the structure function parameters. The figures show the 
68\% CI of the of $\log \mathcal{L}$ evaluated on the log-spaced grid for the sources shown in Fig. \ref{fig:lcfits}. The maximum is marked with a cross, and the values of $\tau_{\mathrm{grid}}$ and $\omega_{r,\mathrm{grid}}$ corresponding to the cross are given in the caption.}}
\label{fig:gridloglik}
 \end{center}
\end{figure*}

One could sensibly derive the pdf's for the parameters $\omega_r,\tau$ and $\vec{\mu}$ via MCMC; however, it proved computationally easier
to calculate $p(m | \omega_r, \tau)$ based on a reasonable parameter grid, and to do the linear optimization of the $\vec{\mu}$ for each grid-point. We use a log-spaced grid of $-2 < \log \omega_r < 0.5$, $0.04 < \tau < 5000$ with 20 values 
evenly spaced in  $\log \omega_r$ and 30 in $\log \tau$ to find the best-fit structure function parameters on the grid $\omega_{r,\mathrm{grid}}$ and $\tau_{\mathrm{grid}}$. We have verified that this approximation agrees well with full MCMC runs.
Fig. \ref{fig:gridloglik} shows the gridded log-likelihood estimates for the same four sources as in Fig. \ref{fig:lcfits}.  The panels show the 68\% CI of the $\log \mathcal{L}$ distribution and the maximum likelihood values of the parameters.

\begin{figure*}[!ht]
              \includegraphics{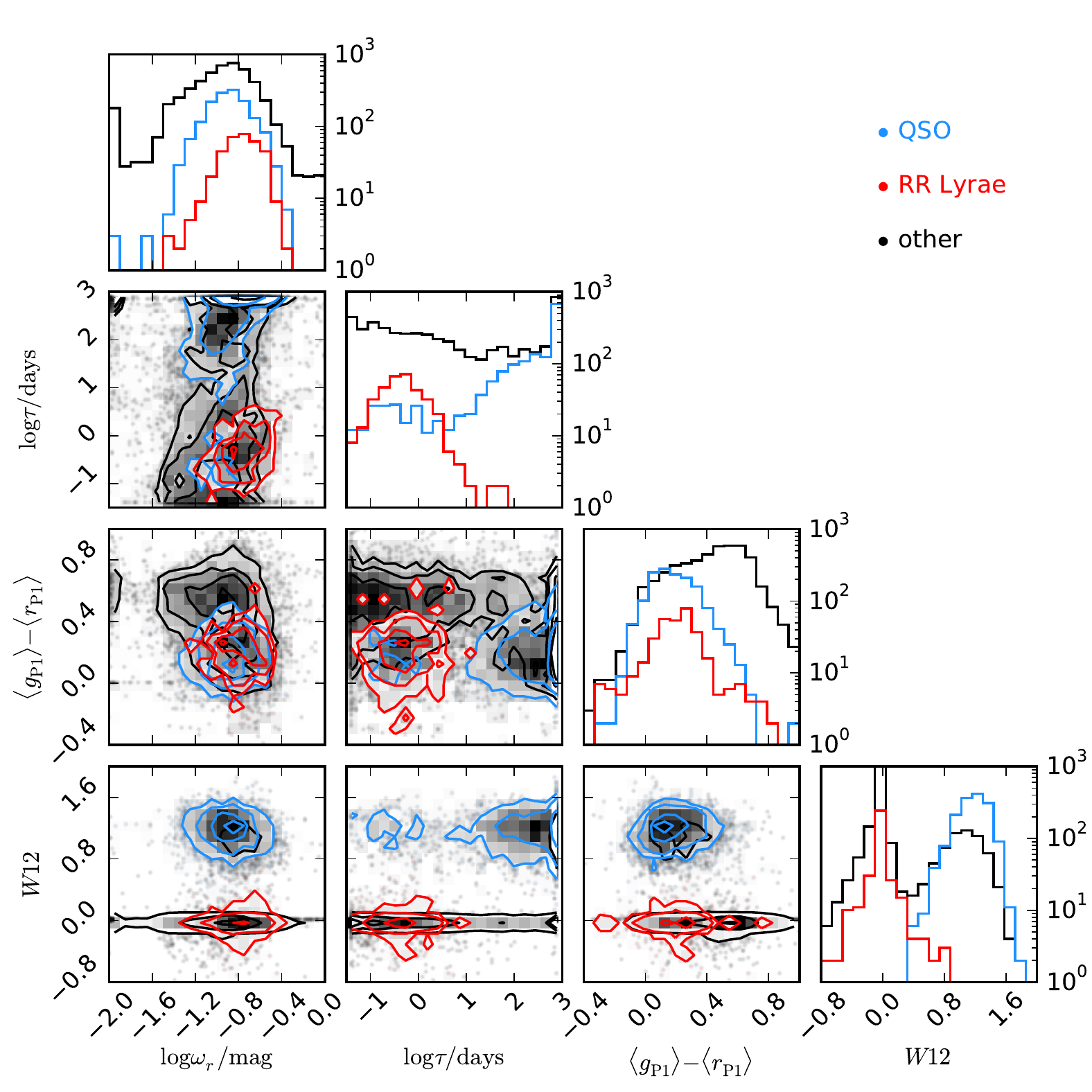}    
                \caption[short]{\footnotesize{Structure function parameters and colors for a subsample of 2380 QSO (blue), 362 RR Lyrae (red), 5196 ``other'' objects (black) surviving magnitude cut and $\hat{\chi}^2>5$, $\hat{\chi}^2>30$ in the stellar locus in S82. Note that for this Figure we calculated the structure function parameter $(\omega_r, \tau)$ using a MCMC, as the discrete griding of $\omega_r$ and $\tau$ proved visually distracting. For $\omega_r$ and $\tau$,
we use the best MCMC point-estimates of the parameters for each source. The $W12$ color (bottom row) illustrates how powerful WISE data are in separating QSOs from other sources \citep{Nikutta2014}.} We presume that most ``other'' sources with $W12>0.5$ are indeed QSOs missed by the SDSS classifiation.}
                                \label{fig:triangleplot}
\end{figure*}

Fig. \ref{fig:triangleplot} shows the distribution of variability parameters $\omega_r$ and $\tau$, for all PS1 objects in the SDSS S82 area that survive the magnitude cut and which have significant variability, either satisfying $\hat{\chi}^2>5$ or $\hat{\chi}^2>30$ for objects within the stellar locus. This Figure illustrates a number of points: first, and unrelated to variability, it shows the power of the WISE color $W1-W2$ to separate QSOs from other sources \citep{Nikutta2014}. Second, it shows that RR Lyrae and QSOs indeed populate different areas of ($\omega_r,\tau$) space.  While they can only be roughly differentiated by their amplitudes $\omega_r$, they have dramatically different time scales $\tau$: RR Lyrae have typical $\tau{\sim} 1$~day and QSOs have $\tau {\sim} 100-1000$ days.

We note that we also explored a power law model for the structure function and found that it provided worse separation between QSO, RR Lyrae and other objects in the structure function parameter plane. This can be explained by the cadence of the survey, as the definition of the power law makes the structure-function fitting more sensitive to the TTI pairs.

Figures \ref{fig:chihatdistribution} and \ref{fig:triangleplot} show that the light curve parameters will be very helpful in classifying variable sources. Yet, these figures also show that simple cuts on some parameters will not be optimal for differentiating object classes. A more sophisticated machine-learning method is needed here.

\subsection{Random Forest Classifier}
\label{sec:RandomForestClassifier}
For classifying objects based on variability measures and mean magnitudes calculated before, we use the Random Forest Classifier (RFC, implemented in Python's \texttt{scikit\_learn} package). Using a training set, it will give the classification probability of a target set's object being of a certain class, $p_{\mathrm{QSO}}$ and $p_{\mathrm{RRLyrae}}$. However, we treat them as arbitrary numbers and not as probabilities, and instead calculate purity and completeness of the sample later on.

For using a RFC, a training set is needed, with observed object parameter values as well as classification labels.
We use the classification of QSOs \citep{Schneider2007,Schmidt2010} and RR Lyrae \citep{Sesar2010} in SDSS S82 as a ground truth.
Positions from SDSS S82 are cross-matched to PS1 positions within 1 arcsec. The object selection and outlier cleaning of Sec. \ref{sec:objectselectionandoutliercleaning} is applied. This results in a set of 7633 QSO, 415 RR Lyrae and more than $1.87 \times 10^6$ other objects. This ``training set template'' is then extended to deal both with magnitude uncertainties and different amounts of reddening.

As a RFC cannot deal with measurement uncertainties by default, we deal with measurement uncertainties by extending the ``training set template'' by copies of itself, sampled within the assumed errors of the PS1 and WISE data. We take 5 samples for each object in the training set, in addition to the original one.

We additionally extend the training set to account for uncertainties
in reddening. Dereddening is done using the reddening-based $E(B-V)$
dust map from \citet{Schlafly2014}.  We extend the training set by
presenting additional QSOs, RR Lyrae, and other objects to the
classifier, where we have artificially introduced a small dereddening
error.  We do this in the following way:
\begin{compactenum}[(i)]
\item make $E(B-V)_{\mathrm{sample}}$ drawn from Gaussian $G (E(B-V)_{\mathrm{catalog}}, \delta E(B-V)=0.1 E(B-V)_{\mathrm{catalog}})$ at the position of the training set source  
\item 5\% chance that $E(B-V)_{\mathrm{sample}}=0$, irrespective of catalog entry
\item sample new mean magnitudes in bands $g_{\rm P1},r_{\rm P1},i_{\rm P1},z_{\rm P1},y_{\rm P1}$ for PS1, and $W1$, $W2$ from WISE within their errors
\item deredden them by $E(B-V)_{\mathrm{sample}}$
\item brighten magnitudes so that $r_{\mathrm{PS1}}$ after dereddening by $E(B-V)_{\mathrm{sample}}$ is the same as after dereddening by $E(B-V)_{\mathrm{catalog}}$.
\end{compactenum}
Each of the sources within the ``training set template'' is re-sampled 5 times to make the training set. This results in a training set having
38165 sources labeled as ``QSO'', and 2075 sources labeled as ``RR Lyrae'' at different reddenings, and a few million the ``other'' objects.

In principle, this technique could be used to train a classifier that
could robustly classify sources even at large $E(B-V)$. In practice,
our current classifier does not however operate reliably at large
$E(B-V)$.  This is because in our training set, we use only objects on
Stripe 82, where the reddening is small, and so no objects with large
reddenings, and, correspondingly, large reddening errors, exist in the
training set.  We defer to later work accurate treatment of highly
reddened objects in the plane, and focus on the lightly reddened
high-latitude regions in this paper.

When using a RFC, missing values have to be replaced by some dummy values (``imputation'') in the training and target sets.
A common solution is replacing missing values by the mean of the available ones. This can be done not only for missing values, but also for unreliable values. As imputation of the median is impractical for the way we process the data, we had tested if an imputation of -9999.99 instead behaves comparably. 
We found that using -9999.99 versus median had no effects on our results.

In addition to imputing missing values, we use imputation when values are considered as unreliable. Accordingly, we impute a value of -9999.99 also in cases where $\sigma_{W1}>0.3$, $\sigma_{W2}>0.3$,  or when magnitude errors are not available.

The Table \ref{tab:RFCparameters} summarizes the parameter set being used for the RFC. 

Though the mean $r$ band magnitude is helpful in detecting RR Lyrae in general, we don't use it here as it introduces a too strong bias in distance, as the training set covers only the range $14.5 \lesssim r_{\rm P1} \lesssim 21.5$ and we want to identify candidates fainter than 20.25 mag.
Among the colors, the dereddened $(i-z)_{\rm P1}$ is a helpful gravity indicator that helps to reduce contamination \citep{Vickers2012}.

\def\arraystretch{1.5}

\begin{deluxetable*}{p{8cm}p{8cm}}
\tabletypesize{\scriptsize}
\tablecaption{Parameter set for the Random Forest Classifier \label{tab:RFCparameters}}
\tablewidth{0pt}
\tablehead{
\colhead{parameter} & \colhead{description} }
\startdata
$\omega_{r,\mathrm{grid}}$,$\tau_{\mathrm{grid}}$	 &	best fit structure function parameter on log-spaced grid \\
$\hat{\chi}^2$ & normalized $\chi^2$ statistic, see Equ. \eqref{eqn:q} \\
$(g-r)_{\rm P1}$, $(r-i)_{\rm P1}$, $(i-z)_{\rm P1}$, $(z-y)_{\rm P1}$ & colors from dereddened PS1 mean magnitudes \\
$ \langle r \rangle_{\mathrm{P1,deredd}}$ & dereddened PS1 mean $r_{\rm P1}$ magnitude, only used for calculation of $p_{\mathrm{QSO}}$ \\
$W12$ & $W1-W2$, helps with QSO identification\\
$i_{\rm P1}-W1$ & separates RR Lyrae from QSO
\enddata
\end{deluxetable*}

\subsection{Verification of the Method Using SDSS S82 Classification Information}
\label{sec:VerificationoftheMethodUsingSDSSS82ClassificationInformation}
In order to test the efficacy of the selection and classification method, we carried out detailed testing on the S82 area, 
using PS1 lightcurves, with the object classifications from S82 \citep{Schneider2007,Sesar2010} as the ``ground truth''.
To quantify purity and completeness of our classifications, we use S82 both as the training and validation set.
A randomly selected 50\% of the $\gtrsim 1.85 \times 10^ 6$ cleaned S82 objects is used for creating the training set,
with the other half as the validation set. 

For any one of the two categories, say RR Lyrae, we can define a candidate sample $\mathcal{S}$ by the choice of a minimum  $p_{\mathrm{RRLyrae}}$. We can then calculate on the basis of the S82 ground truth
the completeness and the purity of this sample. Here, \textit{purity} is defined as the fraction of all RR Lyrae stars in $\mathcal{S}$, and the ``completeness'' is the fraction of actual RR Lyrae stars contained in $\mathcal{S}$.
In both instances, we would expect completeness to be monotonic and purity to be nearly monotonic in $p_{\mathrm{RRLyrae}}$.

For the QSOs, analogous definitions apply. Depending on context, we describe a sample $\mathcal{S}$ either by a cut on $p_{\mathrm{RRLyrae/QSO}}$ , or by the corresponding purity and completeness of this sample as determined on Stripe 82.

\begin{figure*}[H]
\begin{center}    
\subfigure{\includegraphics[trim=0inch 0.2inch 0.0inch 0.2inch, clip=true]{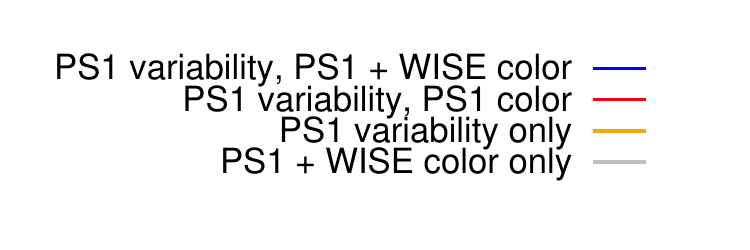}}
    \setcounter{subfigure}{0}
\subfigure[Precision-recall curves showing the trade-off between purity and completeness with regard to total cross-matches sources for different pieces of information provided to the RFC.]{\includegraphics[trim=0inch 0.0inch 0.0inch 0.2inch, clip=true, width=0.75\textwidth]{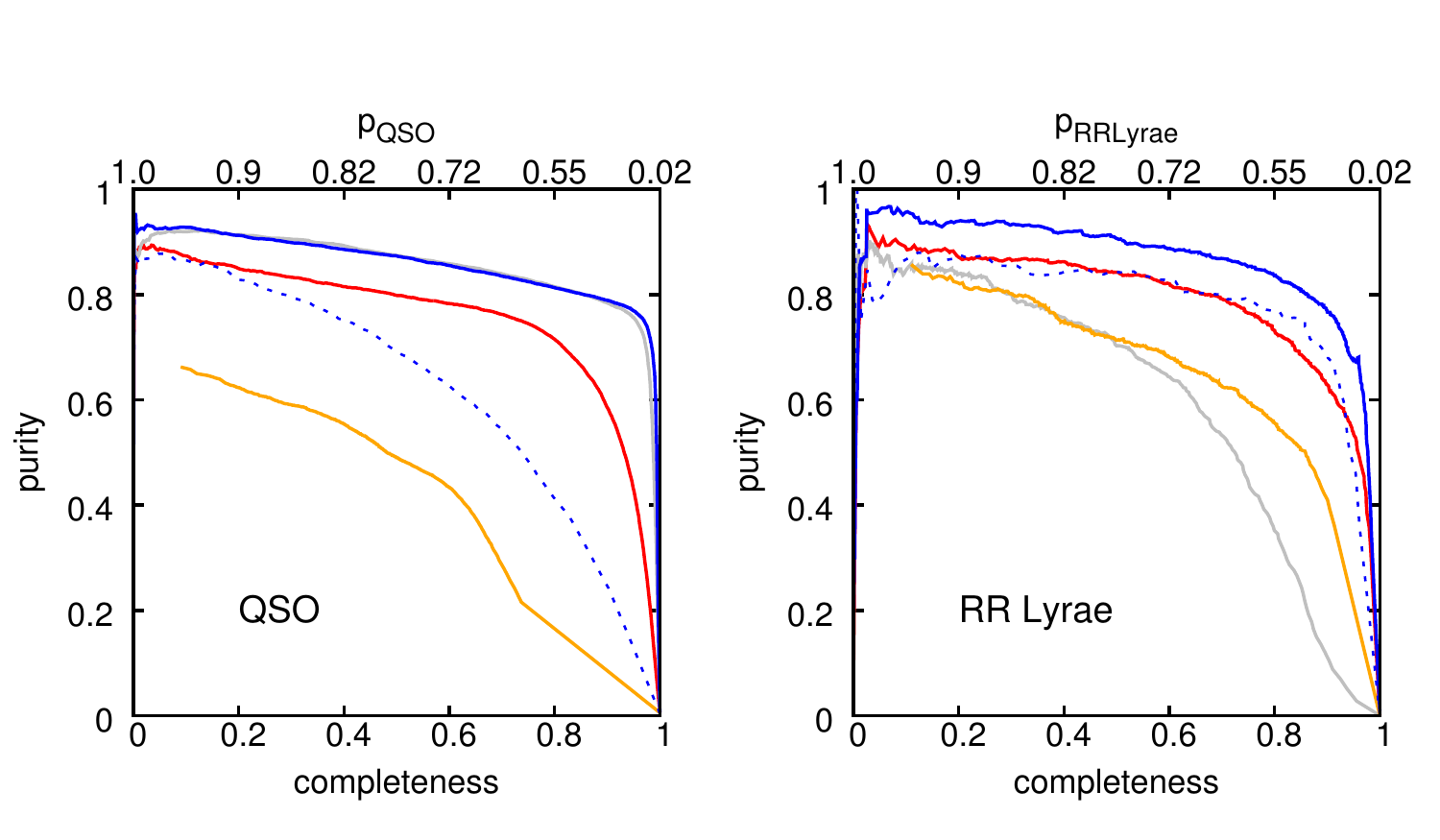}}
\subfigure[Impact of the training-set stochasticisty, illustrated by the dependence of purity and completeness on the chosen training set sources (presuming PS1 variability and PS1 + WISE colors are provided).]{\includegraphics[width=0.75\textwidth]{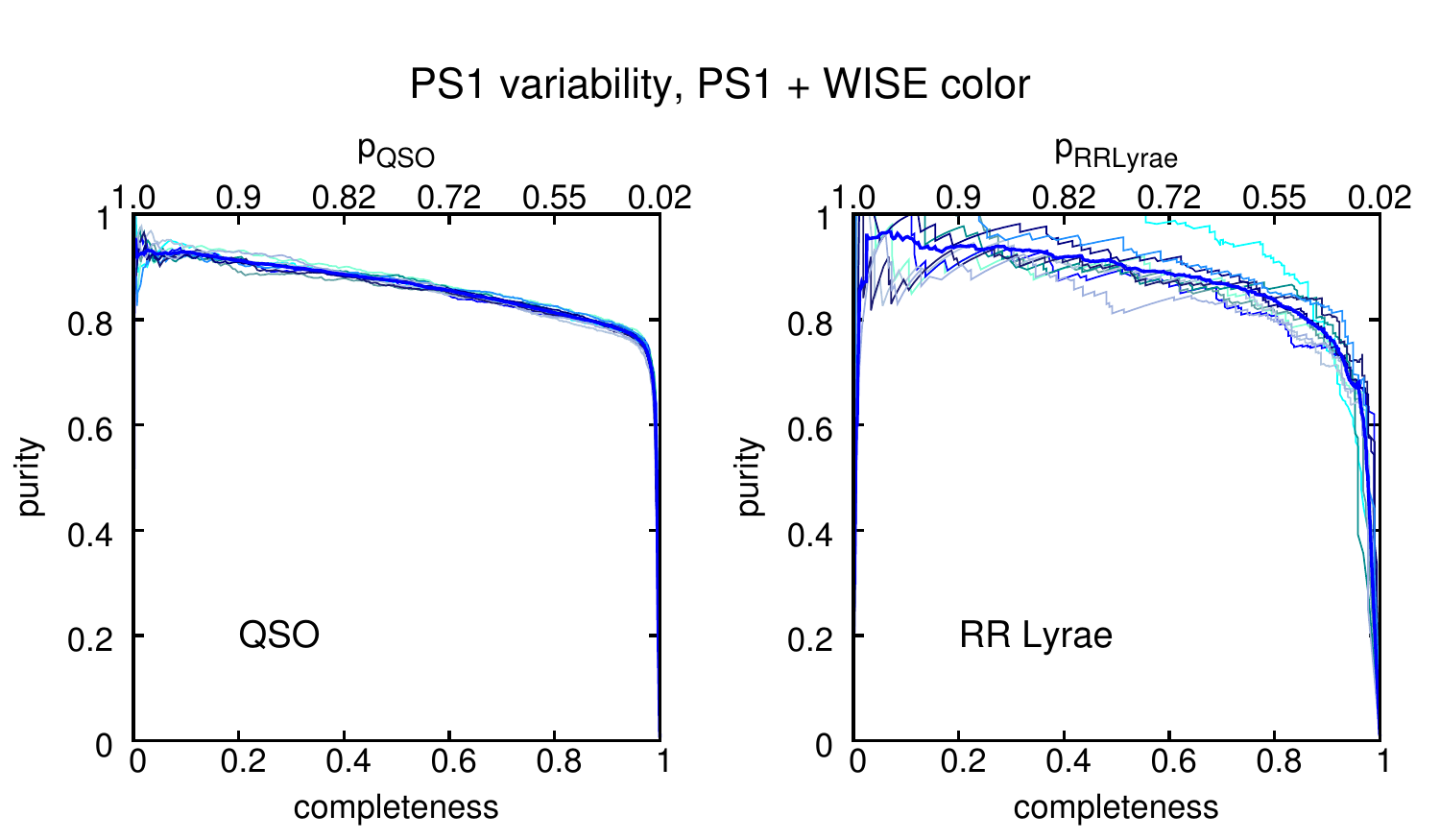}}
\caption{\footnotesize{Trade-off between purity and completeness with respect to total cross-matched sources for different pieces of information provided to the RFC. The upper panels show precision-recall curves when PS1 variability and PS1 + WISE colors, PS1 variability and colors only, PS1 variability only, PS1 + WISE colors are provided. There is a limited purity and completeness that can be achieved with variability only (yellow line).
The numbers for purity and completeness are calculated from bright sources in the S82 training set,
having $15<\langle g_{\mathrm{P1}}\rangle$, $\langle r_{\mathrm{P1}}\rangle$,$\langle i_{\mathrm{P1}}\rangle<20$.
Calculating them instead with respect to all sources ($15<\langle g_{\mathrm{P1}}\rangle$, $\langle r_{\mathrm{P1}}\rangle$,$\langle i_{\mathrm{P1}}\rangle<21.5$) produces the dashed blue lines.
\newline The lower panels show the dependence of purity and completeness on the chosen training set sources when PS1 variability and PS1 + WISE colors are provided (see Tab. \ref{tab:RFCparameters}). The trade-off between purity and completeness is plotted from using 10 different randomly selected training sets, as well as their mean (thick dark blue line). This mean curve is the same as in the upper panel. At the top of the horizontal axis the relation between completeness, and $p_{\mathrm{RRLyrae}}$, $p_{\mathrm{QSO}}$ is given. For RR Lyrae, with only 415 objects in the training set, the stochasticisty is noticeable.}}
\label{fig:spread_puritycompleteness}
 \end{center}
\end{figure*}

Fig. \ref{fig:spread_puritycompleteness} shows precision-recall curves \citep{Powers2011} for the trade-off between purity and completeness with respect to the total cross-matched sources. These values are calculated for sources fulfilling the criteria of having all $\langle g_{\mathrm{P1}}\rangle$, $\langle r_{\mathrm{P1}}\rangle$, $\langle i_{\mathrm{P1}}\rangle$ available and between 15 and 20. We also show the case of all sources ($15<\langle g_{\mathrm{P1}}\rangle$, $\langle r_{\mathrm{P1}}\rangle$, $\langle i_{\mathrm{P1}}\rangle<21.5$) using all available information as dashed blue lines.

The left column refers to QSO classification, the right one to RR Lyrae classification.
This Figure shows that, as expected, for small completeness the purity is maximal, while the completeness is maximized with severe expense to the purity.
What compromise needs to be made  between completeness and purity in sample selection depends in detail on the science question,
but the top panels of  Fig. \ref{fig:spread_puritycompleteness} suggests that the purity increases only little at the expense of completeness less than 80\%. This may be a sensible threshold for an inclusive sample,
whenever PS1 lightcurves and mean colors, as well as  WISE colors are available.
At the top of the horizontal axis we have indicated the relation between completeness and $p_{\mathrm{RRLyrae}}$, $p_{\mathrm{QSO}}$.
Using probability thresholds of ${\sim}0.05$, we get purity both for QSO and RR Lyrae at a level of 70\%, and completeness at a level of 98\%. Using probability thresholds of 0.2, we get purity at a level of 76\%, completeness of 94\% for both QSO and RR Lyrae.

The different lines in the upper panels of Fig. \ref{fig:spread_puritycompleteness} illustrate the relative importance of the different pieces of data that may enter the
 classification; we have not only carried out classification with the
full parameter set from Table \ref{tab:RFCparameters}, but also tested the cases where only
 color-related or variability-related information was used.

For RR Lyrae, Fig. \ref{fig:spread_puritycompleteness} shows that the variability information is absolutely indispensible to define a sample with a interesting combination of purity and completeness. 
For QSOs, (time-averaged) PS1 color together with WISE already do a very good job in selecting QSOs. The PS1 variability provides a significant, but not decisive improvement of purity and completeness. {These different precision-recall curves also indicate what one might expect for purity and completeness, when a particular source lacks some information, for instance, a detection in WISE or particular PS1 bands.}

Given that our training sets are finite in size, the purity and completeness will depend in detail on the chosen training sample.
The individual lines in the lower panels of Fig. \ref{fig:spread_puritycompleteness} reflect different samplings of the training set. For a training set of the size available in S82, the effect is noticeable, but small. 

Completeness and purity may depend on the brightness of objects under consideration. The difference between the blue solid and dashed curves in Fig. \ref{fig:spread_puritycompleteness} show how purity and completeness change when using a bright sample versus using the entire sample.
For QSOs, purity and completeness are significantly reduced, though only a little effect is evident for RR Lyrae. The validity of these conclusions, however, depends on the completeness of our training set at faint magnitudes; see Sec. \ref{sec:LimitationsoftheMethod}.

\begin{figure*}[H]
 \begin{center}  
              \includegraphics[width=0.75\textwidth]{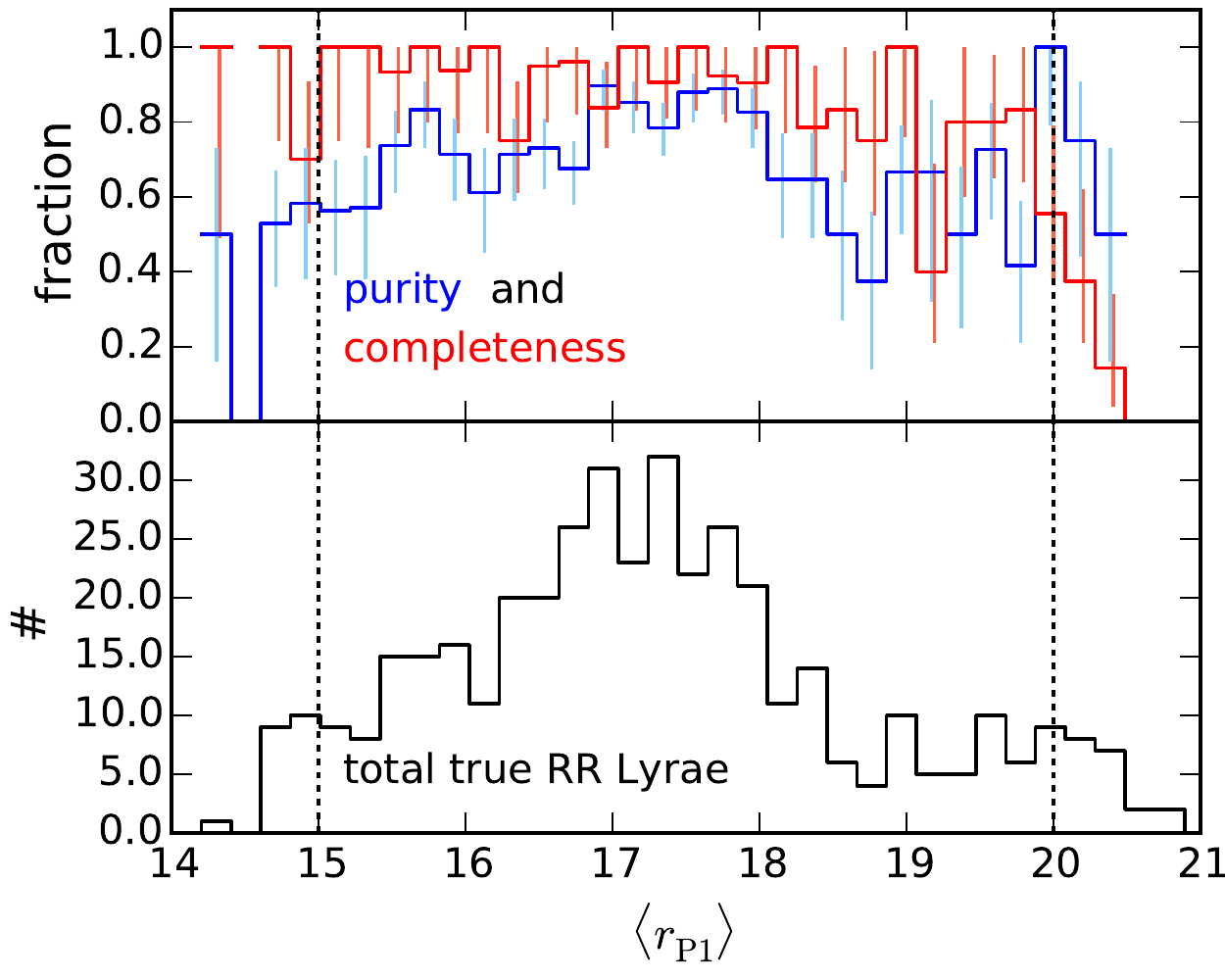}    
                \caption[shortcaption]{\footnotesize{       
Completeness and purity w.r.t. total SDSS S82 RR Lyrae as a function of the mean $r_{\rm P1}$ band magnitude. For a threshold of $p_{\mathrm{RRLyrae}} \geq 0.2$, we get a purity = 75\%, completeness = 92\% within S82. The upper panel gives the dependence of purity and completeness on the apparent $r_{\rm P1}$ band magnitude. The training set consists of the RR Lyrae in the lower panel, showing large variation in the number of sources depending on the mean $r_{\rm P1}$ band magnitude. To account for the different number of sources in different $\langle r_{\rm P1} \rangle$ bins, we calculate error bars on the purity and completeness in the upper panel as the 68\% confidence interval of a Poisson distribution. We find purity and completeness to be roughly constant between 15 and 20 mag. No purity and completeness can be given for $r_{\rm P1}>20.2$, as no object beyond was selected within S82 with a $p_{\mathrm{RRLyrae}} \geq 0.2$.}}
                \label{fig:rrlyrae_completeness_purity_rbandmag}
                 \end{center}
\end{figure*}

To test the performance of the classifier with regard to the signal to noise of the lightcurves, and for RR Lyrae implicitly their distance, we considered the purity and completeness for objects classified through $p_{\mathrm{RRLyrae}} \geq 0.2$, as a function of their apparent magnitude (Fig. \ref{fig:rrlyrae_completeness_purity_rbandmag}). 

For the S82 RR Lyrae sample, using a threshold of $p_{\mathrm{RRLyrae}} \geq 0.2$, we get a purity = 75\%, completeness = 92\% within S82. The upper panel gives the dependence of purity and completeness on the apparent $r_{\rm P1}$ band magnitude. The training set consists of the RR Lyrae in the lower panel, showing large variation in the number of sources depending on the mean $r_{\rm P1}$ band magnitude. To account for the different number of sources in different $\langle r_{\rm P1} \rangle$ bins, we calculate error bars on the purity and completeness in the upper panel from the 68\% confidence interval of a Poisson distribution. 

According to Fig. \ref{fig:rrlyrae_completeness_purity_rbandmag}, we assume the purity and completeness for $p_{\mathrm{RRLyrae}} \geq 0.2$ to be constant within $15 \leq \langle r_{\rm P1} \rangle \leq 20$, but might decrease or be affected by shot noise from the low number of SDSS S82 RR Lyrae beyond this interval. For the detection of faint RR Lyrae, we have to account for a loss in candidates with increasing distance, as discussed in Sec. \ref{sec:DistanceAccuracyFromDracodSph}.

\subsection{Limitations of the Method}
\label{sec:LimitationsoftheMethod}

Our method of automatic source classification is subject to several limitations. The most important of these are:

\begin{compactenum}[(i)]
\item mismatch between our ground-truth training set and other regions of sky,
\item incompleteness of the training set, and
\item the inhomogeneity of the available data over the sky.
\end{compactenum}

We address these limitations in the following subsection.

We train our classifier using data on SDSS S82, where existing large catalogs of RR Lyrae and QSOs provide an almost complete sample of objects in the region.  We wish to train the classifier on this region, but apply it to other regions, where no similar classifications already exist. In general, however, the application of the classifier to regions other than S82 is only justified when the region has distributions of RR Lyrae, QSOs, and potential contaminants similar to that in S82. Over most of the high latitude sky, this is the case, and we can apply our technique without difficulty.

However, at low latitudes the number of contaminants is relatively much larger than the number of RR Lyrae and QSOs in S82, since in these regions the data include very large numbers of metal rich disk stars. Additionally, the data is itself qualitatively different: the presence of reddening changes the colors of sources, and variation in
reddening as a function of distance means that even with a perfect 2D reddening map, dereddened colors may no longer match the true colors of objects. Accordingly, at low latitudes we do not expect our classifier to perform with the same purity and completeness as at high latitudes, and our S82-based estimates of purity and completeness will
no longer apply.

The second problem with our technique is that even in high latitude regions, our adopted training set is imperfect. This is especially
the case for our adopted QSO training set. We use spectroscopically selected QSOs from \citet{Schmidt2010}, which are complete only down to roughly an
$i_{\mathrm{P1}}$-band magnitude limit of 21.25. Therefore, in our training set, fainter objects are marked as non-QSOs, and our classifier learns to
discard these objects -- even when they are, in fact, QSOs, as indicated by their WISE $W1-W2$ color and variability. This results in
a quasar sample from our technique whose purity and completeness is really only relative to S82 spectroscopic quasars, rather than the
underlying population of QSOs falling in our magnitude range.

We expect that our training set is more complete for RR Lyrae, though the very small number of distant RR Lyrae (Figure \ref{fig:rrlyrae_completeness_purity_rbandmag}) means that we
would run the risk there of discarding all distant objects as well, were it not for the fact that we do not include the $r_{\mathrm{P1}}$-band magnitude
as a parameter when classifying RR Lyrae (Table \ref{tab:RFCparameters}).

A final concern with our technique is that our ability to determine if an object is in fact a QSO or RR Lyrae depends on what information is
available for it. The purities and completenesses we compute are properties of the entire sample of selected objects. The assignment
to classes of individual objects within that sample may be relatively uncertain, if, for instance, those objects lack specific PS1 colors or
detections in WISE. Figure \ref{fig:spread_puritycompleteness} serves to show what may happen to the purity and completeness of subsamples of objects, for which only limited information is available.

\section{Results}
\label{sec:Results}

We then applied this variability characterization and subsequent Random Forest classification to all sources in PS1 3$\pi$, with the selection criteria discussed in Section \ref{sec:objectselectionandoutliercleaning}, resulting in a total of more than $3.88 \times 10^ 8$ classified sources. Fig. \ref{fig:healpix_map} shows the all-sky projection of source density within the cuts from Sec. \ref{sec:objectselectionandoutliercleaning}. Here, we present and discuss the results of these classifications. Throughout, we focus in our discussion on two illustrative regimes of Galactic latitude, the North Galactic cap and the Galactic anticenter region. Specifically we selected the regions:
\begin{itemize}
\item $0<l<360$, $60<b<90$ (around the Galactic north pole), about 2800 deg$^2$, about 12 million classified objects    
\item $165<l<195$, $-15<b<15$ (around Galactic anticentre), about 900 deg$^2$, about 20 million classified objects.  
\end{itemize}

As we consider QSOs, but also RR Lyrae, at low galactic latitudes, a number of effects in the candidate selection are likely to become important: first dust extinction at low latitudes will push faint sources below the detection limit; imperfect dereddening may lead to differing de-reddened colors; and the training set, S82, is mostly at high galactic latitudes with low dust, leaving the classifier imperfectly prepared for very high level of Galactic disk star contaminants. 

The large area maps of QSO candidates are shown in Fig. \ref{fig:pqso_patch1} for the North
Galactic cap, in Fig. \ref{fig:pqso_patch2}
for the Galactic anticenter, and in Fig. \ref{fig:mollweide_qso} for the entire PS1 3$\pi$
region. The analogous maps for these three areas, but shown in RR Lyrae candidates are shown on Figures \ref{fig:pRRLyrae_patch1}, \ref{fig:pRRLyrae_patch2} and \ref{fig:mollweide_rrlyrae}.

For both QSOs and RR Lyrae stars these samples of candidates constitute by far the largest sets of high-quality candidates, both in terms of imaging depth, sky area and consequently sample size, e.g. compared to \citet{Morganson2014}, who found a QSO purity of 48\% and completeness of 67\% for PS1-SDSS data. All our candidates are listed in our catalog as described in Sec. \ref{sec:Catalog}.
In the following, all ``purity'' and ``completeness'' given for a threshold on $p_{\mathrm{QSO}}$, $p_{\mathrm{RRLyrae}}$ refer to the case having the full parameter set from Table \ref{tab:RFCparameters} available and making sure the sources fulfilling the criterion of having  all $\langle g_{\mathrm{P1}} \rangle$, $\langle r_{\mathrm{P1}}\rangle$,$\langle i_{\mathrm{P1}}\rangle$ available and between 15 and 20.

\subsection{QSO Candidates}

QSOs should be distributed isotropically across the sky, with a mean number density of candidates, of about 20 objects per deg$^2$ in the magnitude range $15< \mathrm{mag}<21.5$ \citep{Hartwick1990, Schneider2007,Schmidt2010}. This allows us to test the large scale homogeneity of our classification in areas of high Galactic latitude, and it allows us to look at the changing completeness and purity towards low latitudes.
As we expect contaminants to increase at low latitudes, we expect many more candidates with low $p_{\mathrm{QSO}}$. 
Until dust extinction and disk star contamination become severe, we may still get an approximately uniform density of objects with high $p_{\mathrm{QSO}}$.

\begin{figure*}[H]
\begin{center}    
\subfigure[ ]{\includegraphics[width=1\textwidth]{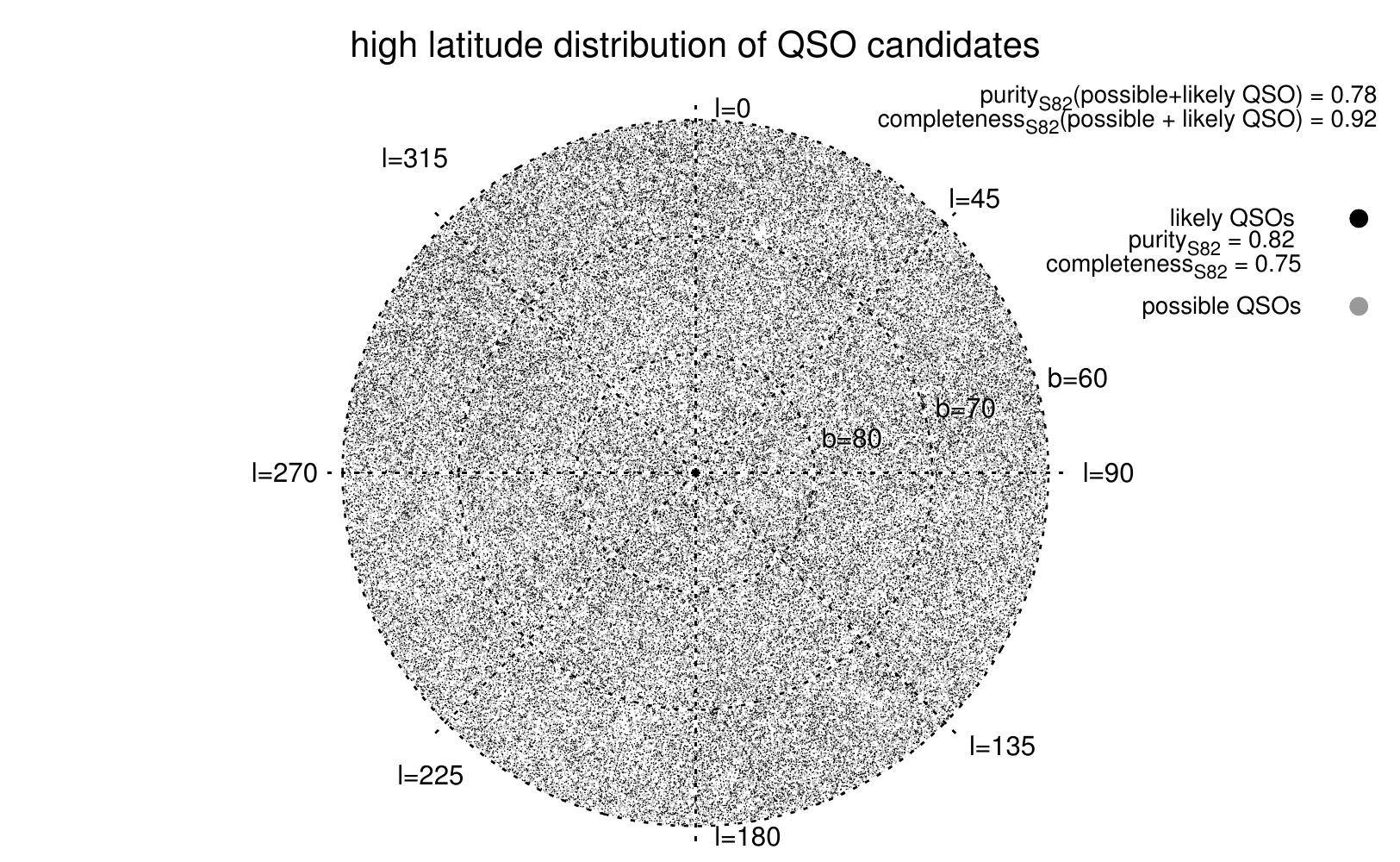}} 
\subfigure[ ]{\includegraphics[width=0.45\textwidth, clip=true]{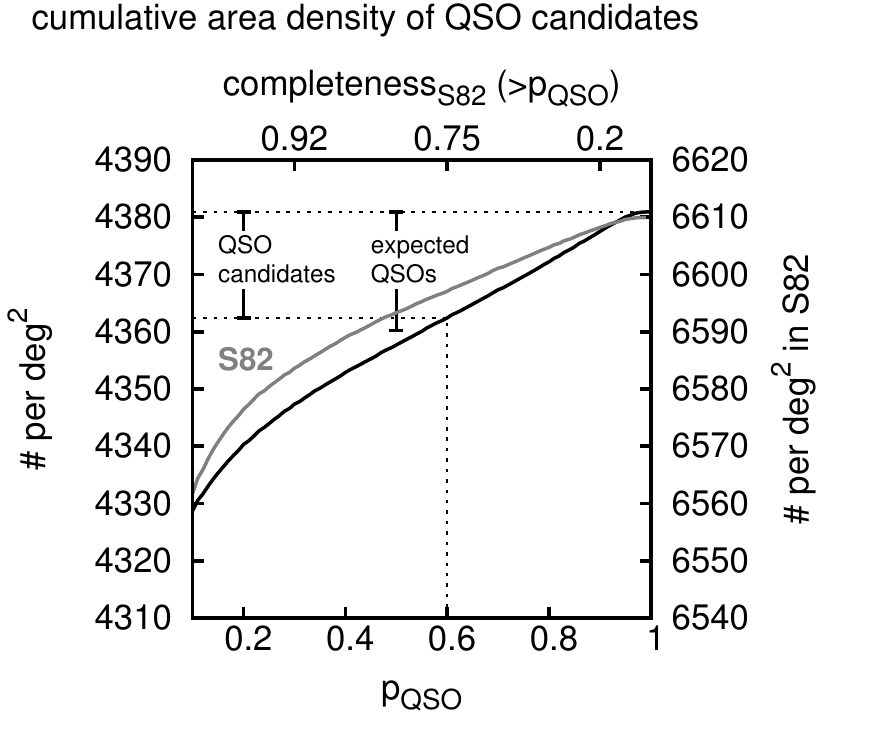}}
\caption{\footnotesize{High latitude distribution of QSO candidates, a) angular distribution of possible and likely QSO candidates showing their uniformity, shown in Lambert's Azimuthal Equal-Area Projection, north polar aspect. b) Cumulative area density of QSO candidates as function of the $p_{\mathrm{QSO}}$ threshold, the vertical lines mark the number of QSO candidates with $p_{\mathrm{QSO}} \geq 0.6$ as well as the expected 20 QSOs per deg$^2$.
At high latitudes, the increase of candidates with $p_{\mathrm{QSO}}$ is similar on and off Stripe 82.}}
\label{fig:pqso_patch1}
 \end{center}
\end{figure*}

Some of these expectations are borne out in the candidate selection near the Galactic North pole: as shown in
Figure \ref{fig:pqso_patch1} the selection of candidates with $p_{\mathrm{QSO}}\ge 0.6$, accounting for a purity in S82 of 82\% and a completeness of 75\%, is uniform to a high degree. 

In regions away from the Galactic plane, we get a homogeneous distribution of the QSO candidates. We see this homogeneity in Fig. \ref{fig:pqso_patch1} as well as in Fig. \ref{fig:mollweide_qso} down to $|b| {\sim}$ 10 deg.
When comparing the increase in the cumulative source density between $p_{\mathrm{QSO}}=0.6$ and $p_{\mathrm{QSO}}=1$, we find an increase of about 20 sources per deg$^2$ (see Fig. \ref{fig:pqso_patch1} (b)).
The number of sources per deg$^2$ at a given minimum $p_{\mathrm{QSO}}$ is compareable for all $|b| > 20\arcdeg$, and compareable to S82. At high latitudes, the increase of candidates with $p_{\mathrm{QSO}}$ is similar on and off S82, as illustrated in Fig. \ref{fig:pqso_patch1}. A sample selected using a lower threshold of $p_{\mathrm{QSO}}$ shows inhomogeneities caused by contamination at almost all Galactic latitudes.

\begin{figure*}[H]
\begin{center}    
\subfigure[ ]{\includegraphics[width=1\textwidth]{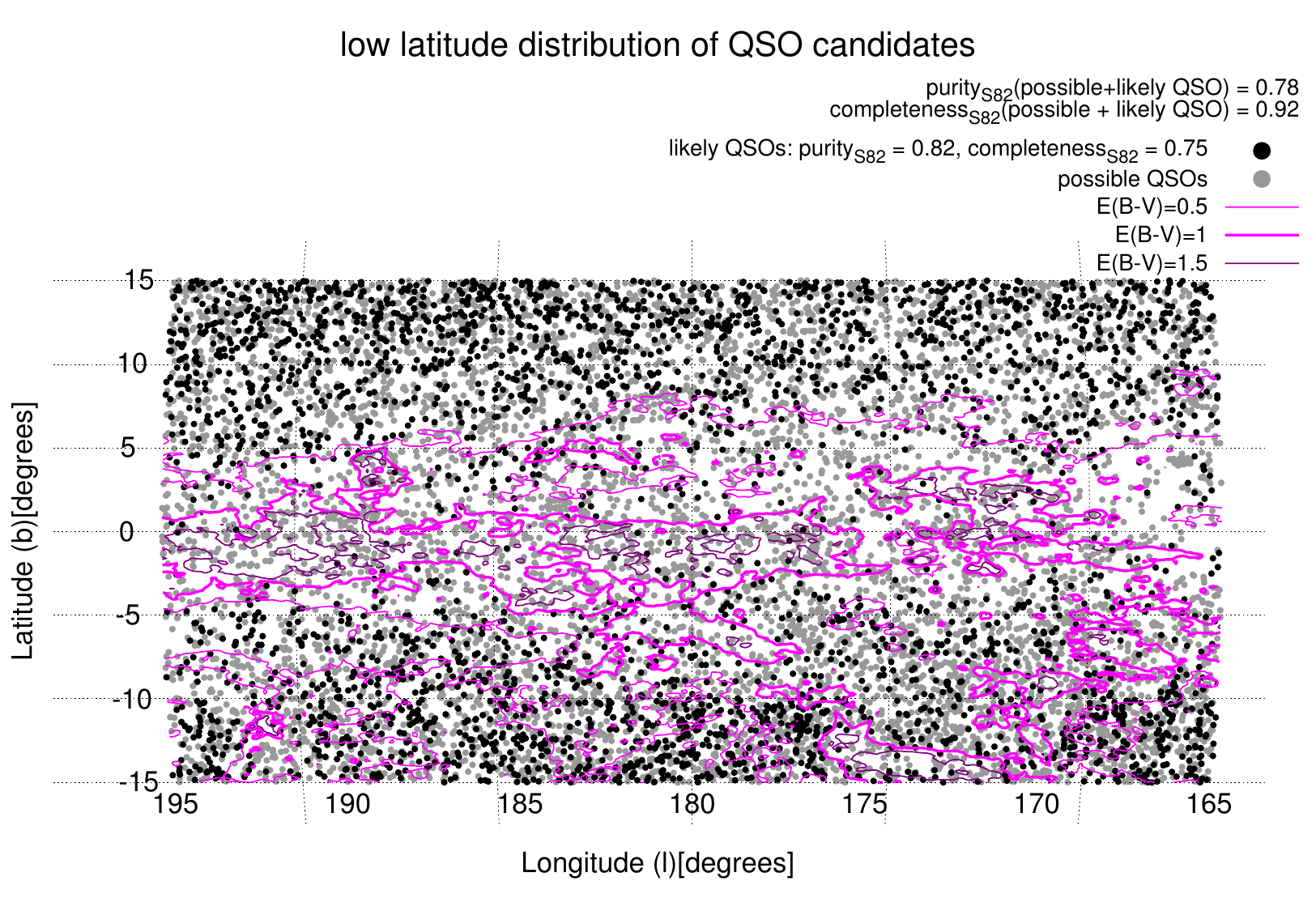}}   
\subfigure[ ]{\includegraphics[width=0.45\textwidth]{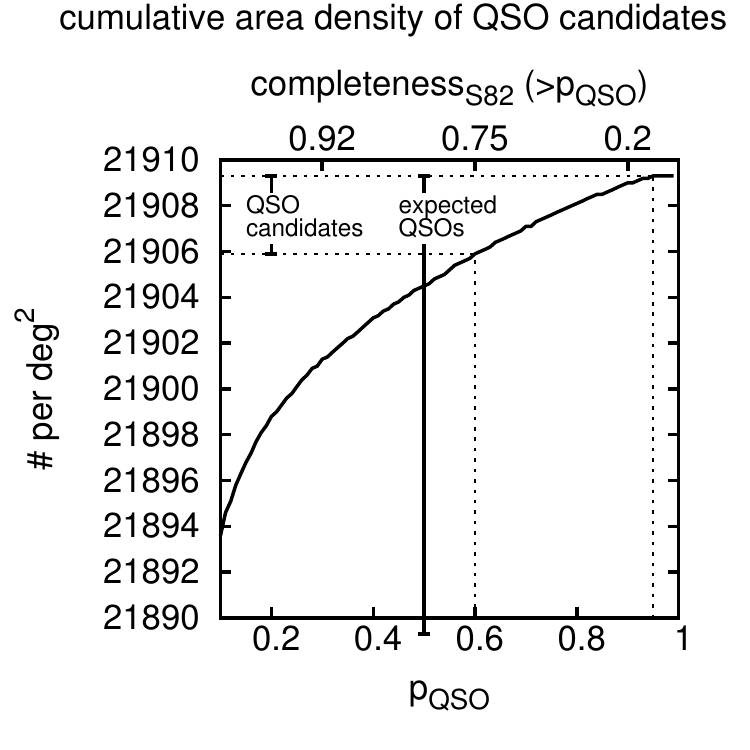}}
\caption{\footnotesize{Low latitude distribution of QSO candidates around the Galactic anticenter, a) angular distribution of possible and likely QSO candidates, shown in Mollweide projection, with a contour plot of the reddening-based $E(B-V)$ dust map \citep{Schlafly2014} overlayed, b) Cumulative area density of QSO candidates as function of the $p_{\mathrm{QSO}}$ threshold, the vertical lines mark the number of QSO candidates with $p_{\mathrm{QSO}} \geq 0.6$ as well as the expected 20 QSOs per deg$^2$. 
Note that at low-latitudes the number of likely QSO candidates ($p_{\mathrm{QSO}}>0.6$), is far below the expectation for an isotropic distribution; dust, varying WISE depth, substantial reddening, and the higher density of contaminants make the secure identification of QSOs more difficult.}}
\label{fig:pqso_patch2}
 \end{center}
\end{figure*}

Around the Galactic anticenter (see Fig. \ref{fig:pqso_patch2}), the number of sources with low $p_{\mathrm{QSO}}$ per deg$^2$
is much higher than around the Galactic north pole, by a factor of ${\sim} 5$. This higher overall source density does not lead to an (presumably erroneous) increase of the number of candidate objects with a high $p_{\mathrm{QSO}}$. Indeed, the number of candidates decreases, caused by dust or varying WISE depth, to less than 2 objects per deg$^2$ (see Fig. \ref{fig:pqso_patch2} (b)).

Across PS1's entire 3$\pi$ area, we find 399,132 likely QSO candidates with $p_{\mathrm{QSO}} \geq 0.6$ (with an expected high-latitude purity=82\%, completeness=75\%), 892,131 candidates with $p_{\mathrm{QSO}} \geq 0.2$ (purity =77\%, completeness=95\%)  and 1,596,319 possible candidates (purity = 0.72\%, completeness= 98\%) with $p_{\mathrm{QSO}} \geq 0.05$.

\subsection{RR Lyrae Candidates}

In this section, we present the properties of the resulting RR Lyrae candidate sample. In particular, we test whether the completeness and purity of our selection is good enough to recover known halo substructure, as well as whether it can compete with the classification from other surveys the method is not trained on.

\begin{figure*}[H]
\begin{center}    
\subfigure[ ]{\includegraphics[width=1\textwidth]{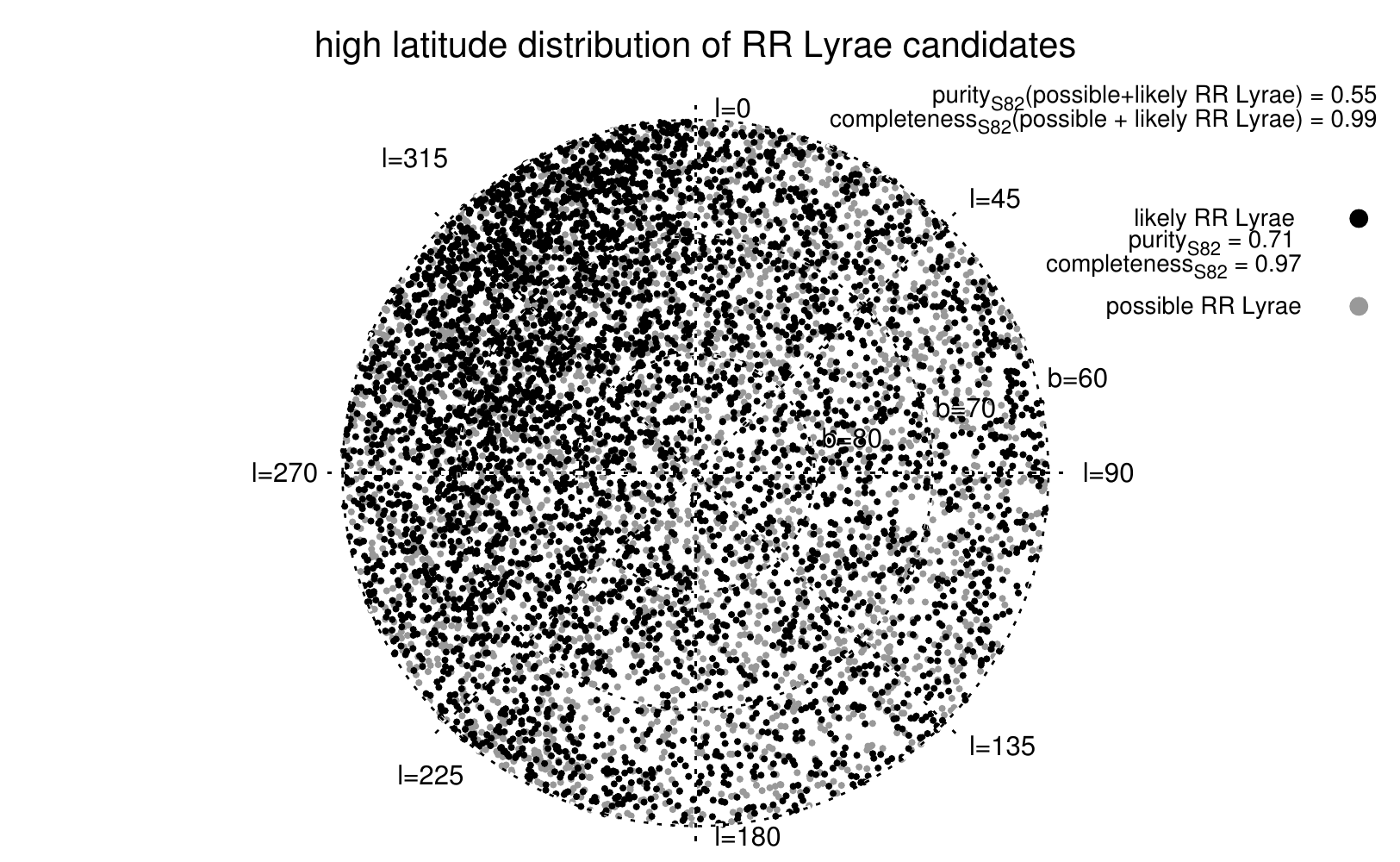}} 
\subfigure[ ]{\includegraphics[width=0.45\textwidth, clip=true]{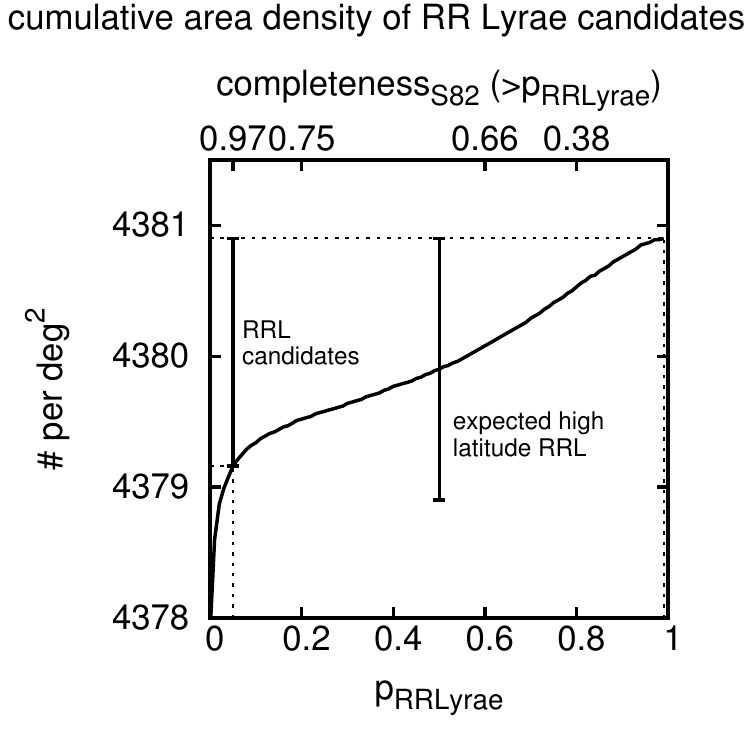}}
\caption{\footnotesize{Distribution of $p_{\mathrm{RRLyrae}}$ around Galactic north pole, a) distribution of likely contaminants ($p_{\mathrm{RRLyrae}}<0.05$) and RR Lyrae candidates ($0.05 \leq p_{\mathrm{RRLyrae}} \leq 1$) across the area, shown in Lambert's Azimuthal Equal-Area Projection, north polar aspect, b) Cumulative area density of RR Lyrae candidates as function of the $p_{\mathrm{RRLyrae}}$ threshold, the vertical lines mark the number of RR Lyrae candidates with $p_{\mathrm{RRLyrae}} \geq 0.05$ as well as the expected average of 2 high latitude RR Lyrae per deg$^2$.}}

\label{fig:pRRLyrae_patch1}
 \end{center}
\end{figure*}

\begin{figure*}[H]
\begin{center}    
\subfigure[ ]{\includegraphics[width=1.0\textwidth]{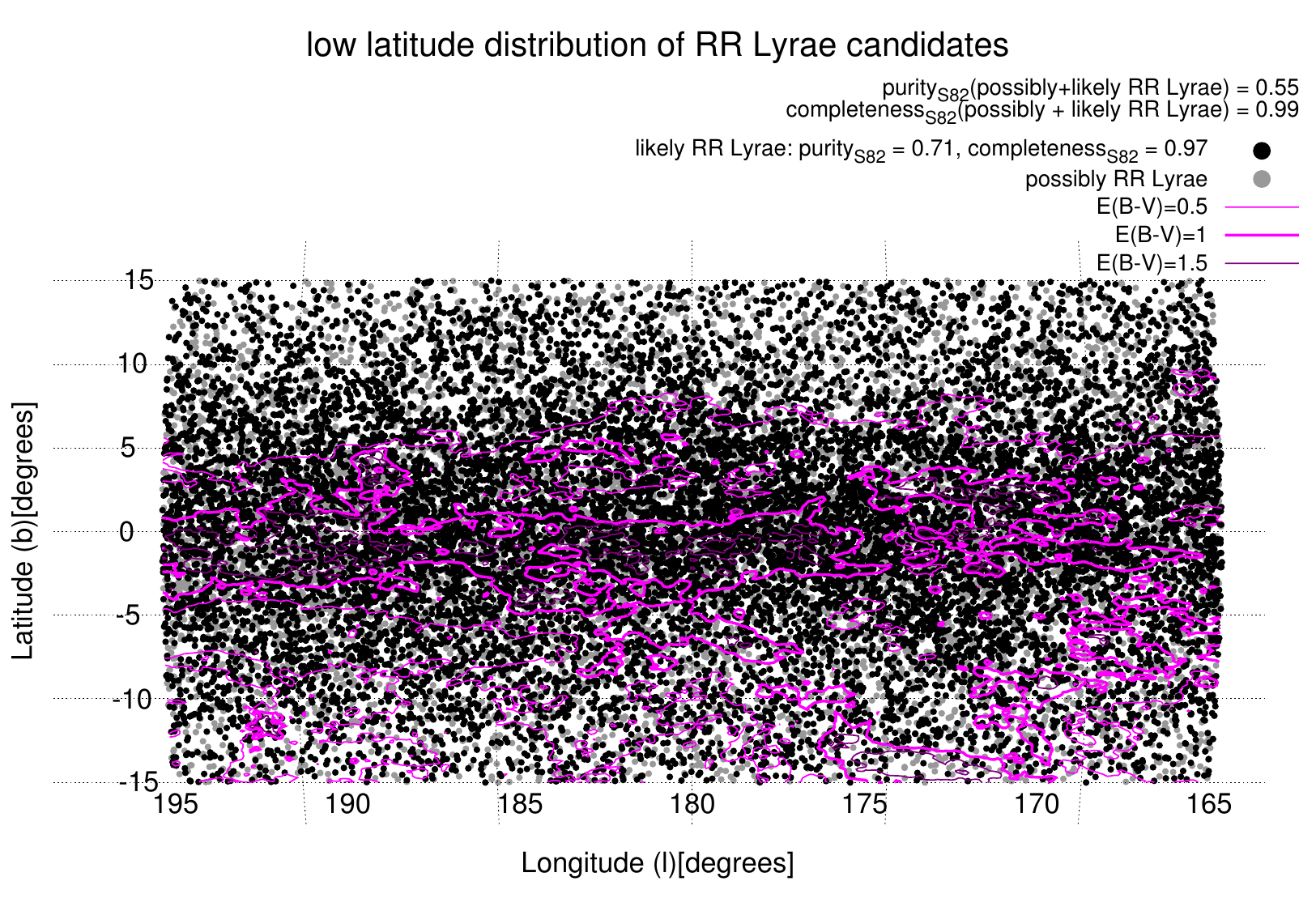}} 
\subfigure[ ]{\includegraphics[width=0.45\textwidth]{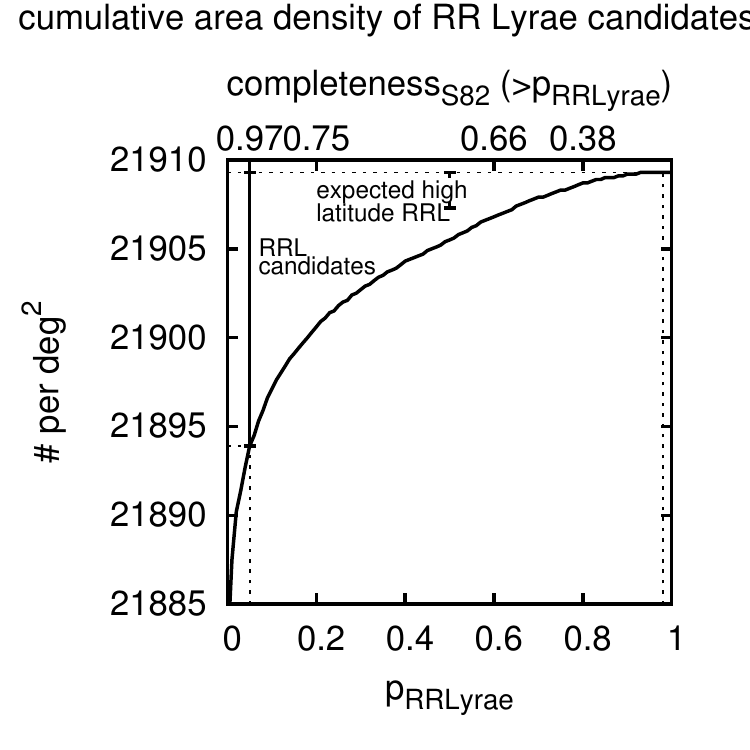}}   
\caption{\footnotesize{Distribution of $p_{\mathrm{RRLyrae}}$ around Galactic anticenter,  a) distribution of likely contaminants ($p_{\mathrm{RRLyrae}}<0.05$) and RR Lyrae candidates ($0.05 \leq p_{\mathrm{RRLyrae}} \leq 1$) across the area, Mollweide projection, b) Cumulative area density of RR Lyrae candidates as function of the $p_{\mathrm{RRLyrae}}$ threshold, the vertical lines mark the number of RR Lyrae candidates with $p_{\mathrm{RRLyrae}} \geq 0.05$ as well as the expected average of 2 high latitude RR Lyrae per deg$^2$. The wide discrepancy between the number of RR Lyrae candidates to those expected from high latitude is a combination of presumably true ``disk RR Lyrae'' and higher contamination.}}
\label{fig:pRRLyrae_patch2}
 \end{center}
\end{figure*}

Figures \ref{fig:pRRLyrae_patch1}, \ref{fig:pRRLyrae_patch2} and \ref{fig:mollweide_rrlyrae} present the diagnostics of our RR Lyrae candidate identification, analogous to the Figures for the QSO candidates. Because we expect the angular and 3D distribution of RR Lyrae to be highly structured, diagnosing the quality of our candidate identification across PS1 3$\pi$ is more complex than for the QSOs.
Even Figure \ref{fig:pRRLyrae_patch1}, showing the  RR Lyrae distribution around the Galactic north pole in its top-panel, 
shows gradients and structure; the overdensity seen between $l=220\arcdeg$ and $315\arcdeg$ is the Sagittarius (Sgr) tidal stream. The bottom panel of Fig. \ref{fig:pRRLyrae_patch1} shows that we have one very likely RR Lyrae 
($p_{\mathrm{RRLyrae}} \geq 0.6$, purity=88\%, completeness=66\%) per deg$^2$ and two per deg$^2$ with $p_{\mathrm{RRLyrae}} \geq 0.05$ (purity=71\%, completeness=98\%). This fits with the expectation of about 1-2 RR Lyrae per deg$^2$ from SDSS S82 \citep{Sesar2010}.

At low latitudes, around the Galactic anticenter (see Fig. \ref{fig:pRRLyrae_patch2}) where the total source density is 5 times higher,
the number of RR Lyrae candidates with $p_{\mathrm{RRLyrae}} \geq 0.6$ is comparable, only 1--2/deg$^2$. This may reflect the combination of higher contamination (reducing the number of $p_{\mathrm{RRLyrae}} \geq 0.6$ candidates), with actual RR Lyrae in the Galactic disk. The number of possible RR Lyrae (candidates), with $p_{\mathrm{RRLyrae}} \geq 0.05$  is much higher than around the Galactic north pole, by a factor of ${\sim} 5-10$, which must reflect, foremost, increased contamination. Compared to the QSO's, we have chosen a more inclusive criterion for further consideration of RR Lyrae candidates, i.e. $p_{\mathrm{RRLyrae}} \geq 0.05$, because subsequent period-fitting (Sesar et al in prep.) can dramatically reduce the contamination, while preserving high completeness. 

The panoptic view of the PS1-selected RR Lyrae candidates (Fig. \ref{fig:mollweide_rrlyrae}) is quite striking, as it reveals how prominent the Galactic disk and bulge are in the map of likely RR Lyrae candidates. Note that this is in stark contrast to the large-scale distribution of probable QSOs, whose density drops towards the Galactic plane. Therefore, these data may, in addition to contaminants, be revealing enormous numbers of RR Lyrae candidates throughout the disk and the bulge. Bulge RR Lyrae have been surveyed extensively, e.g. by OGLE \citep{Udalski2003}; yet, to date there have been very few studies of RR Lyrae throughout the Galactic disk \citep{Mateau2012}. This survey therefore represents the largest sample of Galactic disk RR Lyrae candidates, by a wide margin. Of course, they require extensive verification and follow-up.

We find 59,888 possible candidates with $p_{\mathrm{RRLyrae}} \geq 0.05$ (purity=71\%, completeness=97\%) and 42,674 highly likely halo RR Lyrae candidates at Galactic latitudes of $|b| >  20\arcdeg$ outside of the bulge, having $p_{\mathrm{RRLyrae}} \geq 0.2$ (purity=75\%, completeness=92\%).

Within $|b| < 20\arcdeg$, where reddening and contamination mean our method is less likely to be reliable (Section \ref{sec:Methodology}),  we find 187,393 possible RR Lyrae candidates with $p_{\mathrm{RRLyrae}} \geq 0.05$ and 
110,474 highly likely RR Lyrae candidates with $p_{\mathrm{RRLyrae}} \geq 0.2$. Out of them, 19,958 with $p_{\mathrm{RRLyrae}} \geq 0.05$ and 12,967 with $p_{\mathrm{RRLyrae}} \geq 0.2$ belong to the bulge as being in a radius of 20$\arcdeg$ around the Galactic center. From this, we get 167,435 possible and 97,510 highly likely disk RR Lyrae candidates outside of the bulge.
Within the complete area covered by PS1 3$\pi$, we find 247,281 possible RR Lyrae candidates and 153,151 highly likely RR Lyrae candidates.

At higher Galactic latitudes, the PS1 3$\pi$ includes sky regions with known halo substructures or satellite galaxies that contain RR Lyrae, and we can make use of this to verify our candidate selection. Known structures, clusters and satellite galaxies are labelled\footnote{ http://homepages.rpi.edu/~newbeh/ newline mwstructure/MilkyWaySpheroidSubstructure.html} in Fig. \ref{fig:mollweide_rrlyrae}. Many of them show up in the our map of likely RR Lyrae. Note that M31 and M33 appear in these maps, presumably because their (unreddened) Cepheids get misintepreted as RR Lyrae by our classifier.

\subsubsection{The Sagittarius Stream}

\begin{figure*}[!ht]       
  \begin{center} 
\subfigure[ ] {\includegraphics[width=0.66\textwidth, trim=0inch 0.34inch 0.0inch 0.13inch, clip=true]{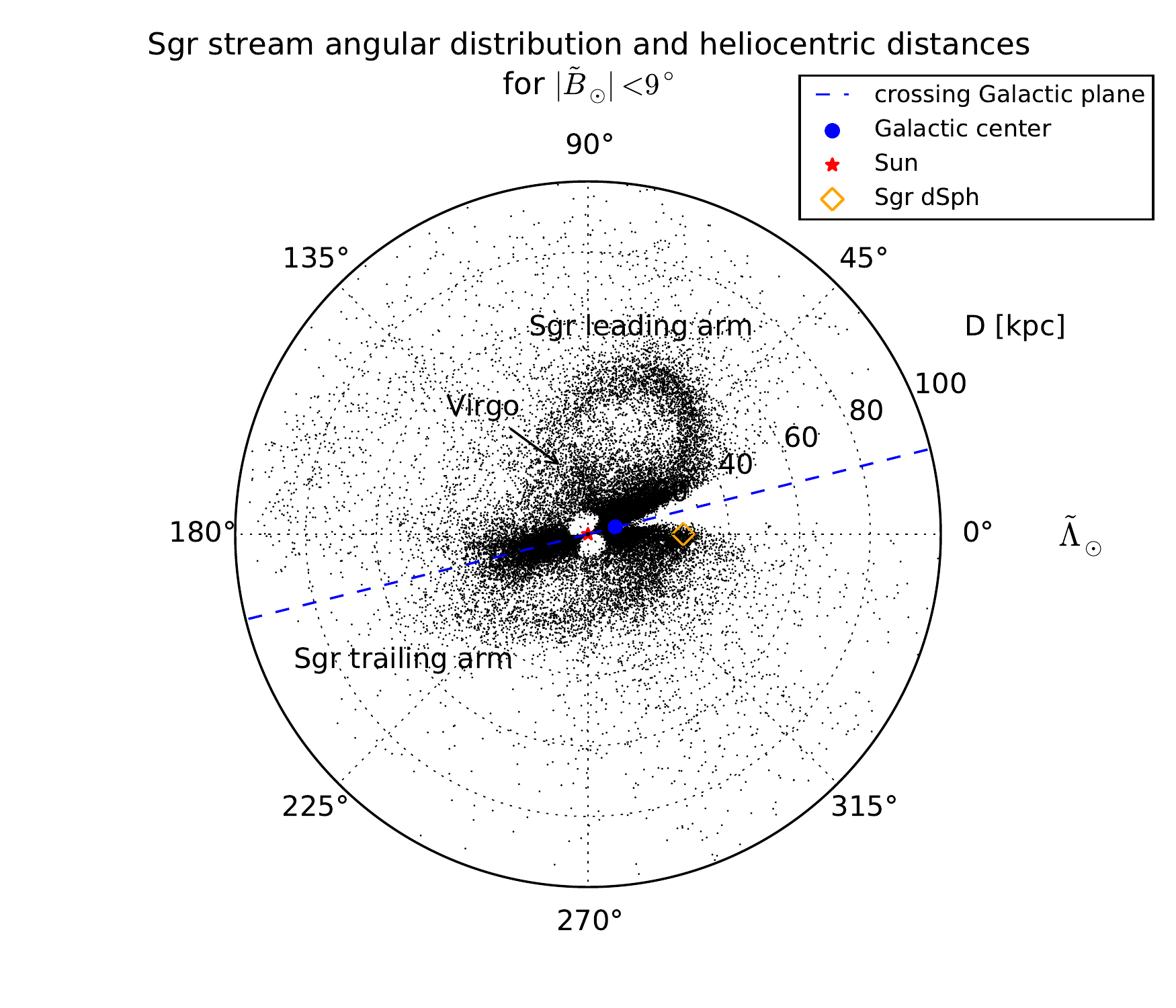} }
 \subfigure[ ] {\includegraphics[width=0.66\textwidth, trim=0.1inch 0.0inch 0.0inch 0.11inch, clip=true]{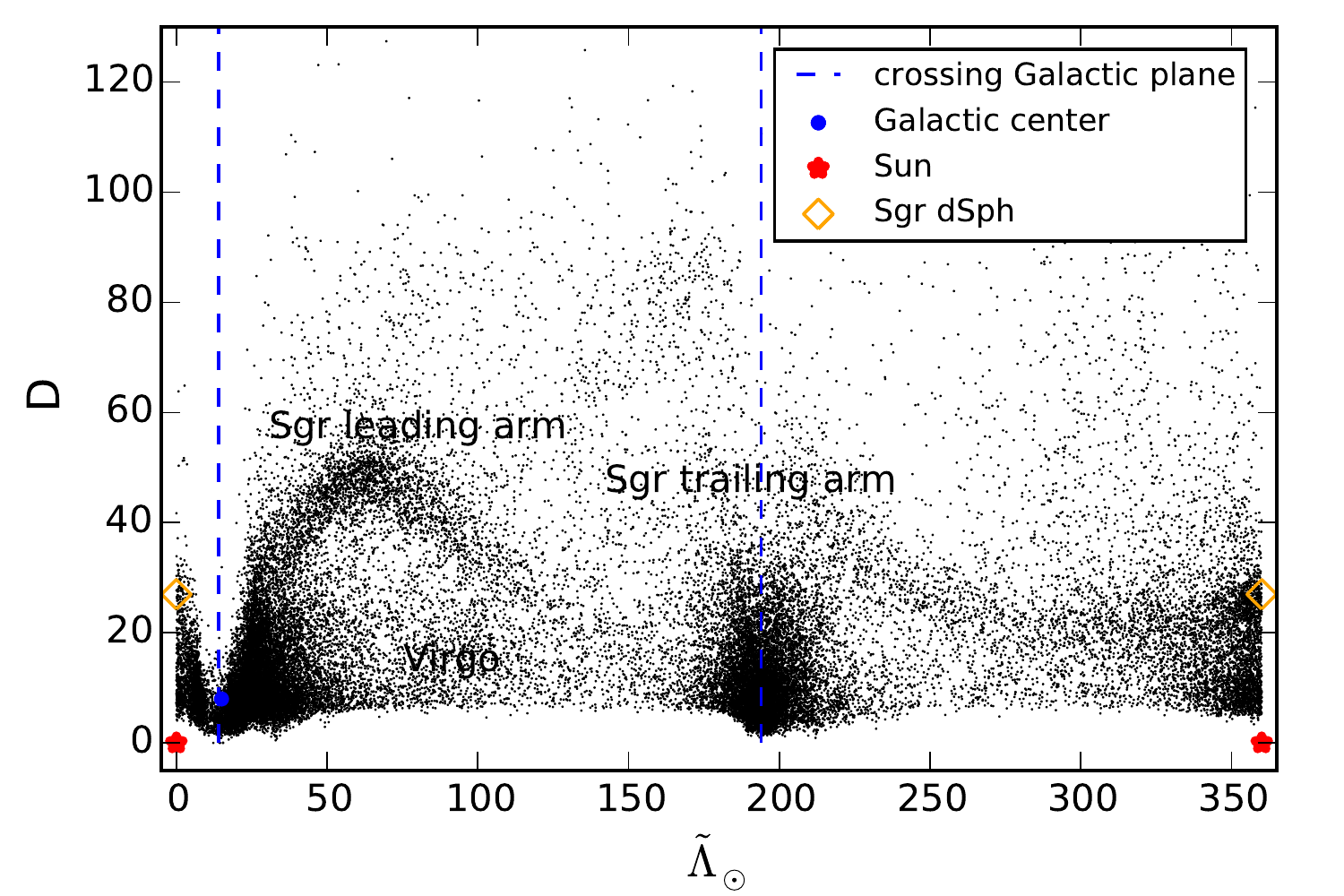}}
                \caption{\footnotesize{
                The extent of the Sagittarius tidal stream from the distribution of RR Lyrae candidates ($p_{\mathrm{RRLyrae}}  \geq 0.05$, purity=71\%, completeness=97\%). The leading and trailing arm of Sagittarius stream can be identified, as well as several substructures up to more than 100 kpc. Distances are from distance modulus of dereddened $r$ band mean magnitude. The longitudes of the crossing Galactic plane at $l=14\arcdeg$ and $l=190\arcdeg$ are marked. \newline
                (a) Distribution of RR Lyrae candidates ($p_{\mathrm{RRLyrae}}  \geq 0.05$) within $\pm$9 deg of the Sagittarius plane, shown in Sagittarius coordinates from \cite{Belokurov2014}, \newline (b) Alternative projection of the RR Lyrae candidates in the Sagittarius tidal stream.}}
                \label{fig:sgr_coords_belokurov_r}
                \end{center}
\end{figure*}

The dominant substructure in the Galactic halo (aside from the Magellanic clouds) is the Sagittarius stellar stream, with its leading and trailing arms 
\citep{Majewski2003}. 
Already in Figure \ref{fig:mollweide_rrlyrae}, the Sagittarius stellar stream can be seen as an overdensity crossing $l=0\arcdeg$ and $l=180\arcdeg$. It is useful to show the geometry of the Sagittarius stream by selecting stars near its presumed orbital plane, and then showing a projected view of this orbital plane. 
Specifically, we plot the angular and distance distribution for RR Lyrae candidates with $p_{\mathrm{RRLyrae}} \geq 0.2$ (formal purity=76\%, completeness=94\%) in Fig. \ref{fig:sgr_coords_belokurov_r} using the heliocentric Sagittarius (orbital plane) coordinates $(\tilde{\Lambda}_{\odot}, \tilde{B}_{\odot})$ defined by \cite{Belokurov2014} and a distance modulus $D$ from the mean magnitude $\langle r \rangle$.
In this coordinate system, the equator is aligned with the plane of the Sagittarius trailing tail, and $\tilde{\Lambda}_{\odot}$ increases in the direction of Sagittarius motion. The latitude axis $\tilde{B}_{\odot}$ points to the Galactic North pole.

Distances $D$ in parsec were taken from
\begin{equation}
D = 10^  {\left( \left( \langle r\rangle_{\mathrm{deredd}} -M_r+5\right) /5 \right)}.
\end{equation}
where $\langle r\rangle_{\mathrm{deredd}}$ is the dereddened $r$ mean magnitude.
We adopt $M_r {\sim} M_V$=0.60 mag from \citet{Sesar2010} who used the \citet{Chaboyer1999} $M_V-\mathrm{[Fe/H]}$ relation under the assumption that the mean metallicity of RRab stars is equal to the median metallicity of halo stars \citep[$\mathrm{[Fe/H]=-1.5}$,][]{Ivezic2008}. As we don't distinguish between RRab and RRc stars from our analysis, and RRab stars are most common, making up 91\% of the observed RR Lyrae, we use $M_r {\sim} M_V =0.60$ mag for all RR Lyrae candidates.

Figure \ref{fig:sgr_coords_belokurov_r}, showing the RR Lyrae candidates in the Sagittarius plane, provides a striking view of the stream, with its trailing and leading arm to distances of about 100 kpc.
We can compare the structure in Fig. \ref{fig:sgr_coords_belokurov_r} to Fig. 6 in \citet{Belokurov2014} as well as to Fig. 6 and 17 in \citet{Law2010} that shows the best-fit N-body debris model in a triaxial halo and observational constraints from 2MASS + SDSS for the leading and trailing arm.

We compare our results to \citet{Ruhland2011}, who traced the Sagittarius stellar stream using BHB stars and compared it to \citet{Law2005}. From our results, we can
confirm that there is an extension of the trailing arm at distances of 60 -- 80 kpc from the Sun as given e.g. by \citet{Ruhland2011}.
Furthermore, we find a cloud-like overdensity at $\tilde{\Lambda}_{\odot} {\sim} 110 \arcdeg$, $5 \lesssim D\lesssim 25$ kpc, that can be identified with the Virgo overdensity.
This overdensity can be seen in a number of works \citep{Ruhland2011, Cole2008, Newberg2007}, but our RR Lyrae candidates show the three-dimensional structure especially clearly.


\subsubsection{Distance Accuracy from the Draco dSph}
\label{sec:DistanceAccuracyFromDracodSph}
The Draco dwarf galaxy, at known distance and known to contain many RR Lyrae, provides us with an opportunity to estimate the distance precision of the RR Lyrae candidates, using their inferred mean magnitude in the $r$ band. As we expect many RR Lyrae in this direction, we consider an inclusive set of candidates with $p_{\mathrm{RRLyrae}} \geq 0.05$. Draco is entirely dominated by old stars, and is affected by near-negligible 
reddening, which increases the likelihood of dealing with true RR Lyrae 
stars as compared to the candidates seen in the region of the Galactic disk.
Out of the 272 RR Lyrae listed by \citet{Kinemuchi2008}, in 248 cases we found a cross-matching PS1 source within our cuts, which compares well with our 10\% selection loss (Section \ref{sec:objectselectionandoutliercleaning}). 

\begin{figure*}[!ht]
 \begin{center}  
\subfigure[ ]{\includegraphics[width=0.45\textwidth]{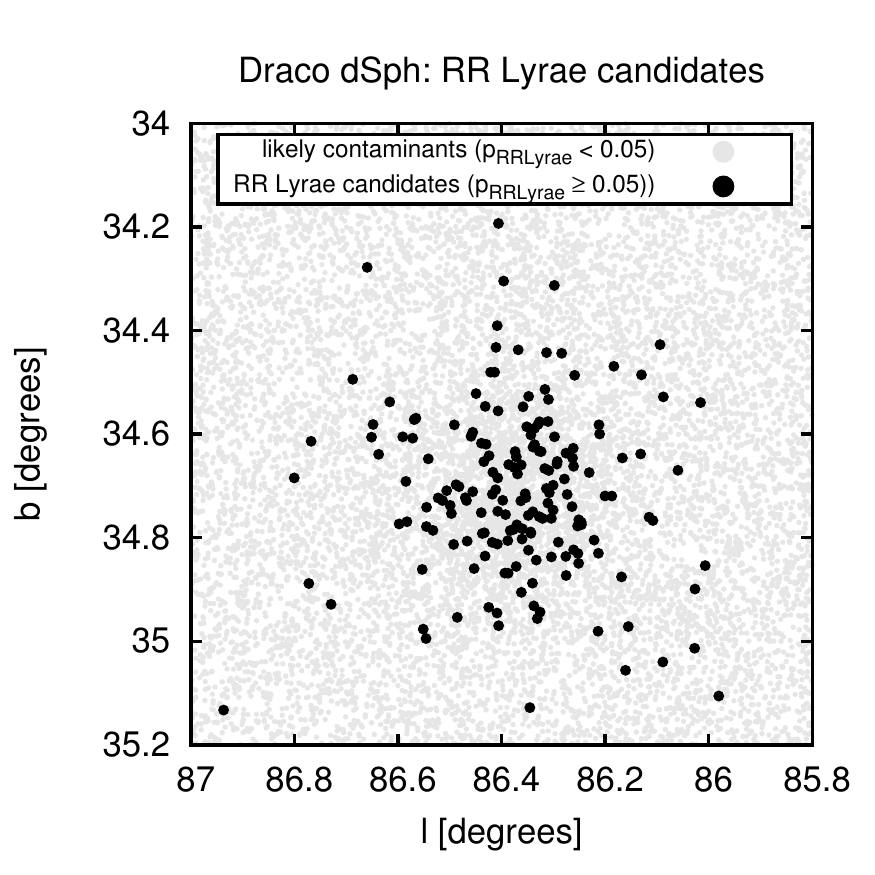}}
\subfigure[ ]{\includegraphics[width=0.45\textwidth]{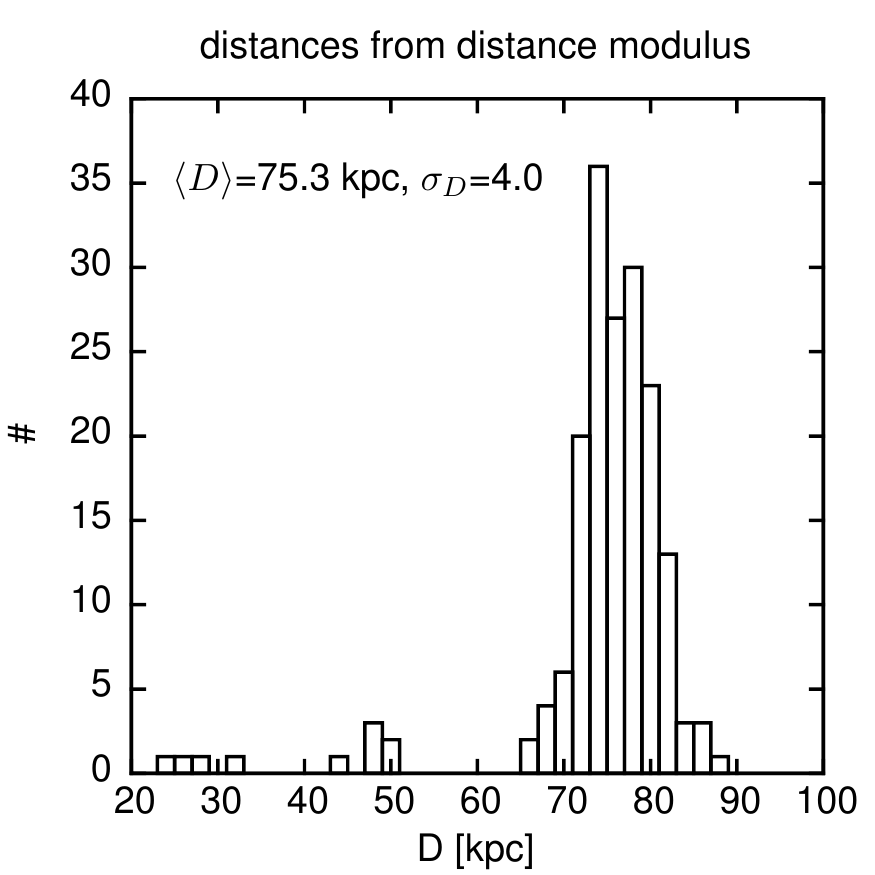}}
                                \caption[]{\footnotesize{Illustration of the distance precision for RR Lyrae candidates ($p_{\mathrm{RRLyrae}} \geq 0.05$) shown in their distribution around Draco dSph. (a) angular distribution, compared to that of likely contaminants, (b) distance estimates from distance modulus of dereddened $r_{\rm P1}$ band mean magnitude for RR Lyrae candidates ($p_{\mathrm{RRLyrae}} \geq 0.05$). The distance estimates are in good agreement with \citet{Kinemuchi2008} and \citet{Bonanos2004}.} Because Draco's RR Lyrae are fainter than most RR Lyrae in the training set, the completeness with respect to the sources found by \citet{Kinemuchi2008} is only 61\%.} 
                \label{fig:draco}
                 \end{center}
\end{figure*}

Out of these, 164 have a $p_{\mathrm{RRLyrae}} \geq 0.05$. The first panel of Fig. \ref{fig:draco} shows their angular distribution (black points); the second panel shows their distribution in distance $D$. Our result of $75.3 \pm 4.0$ kpc is in good agreement with \citet{Kinemuchi2008} who found a distance of $82.4 \pm 5.8$ kpc, and \citet{Bonanos2004} who found a distance of $75.8 \pm 5.4$ kpc. Remarkably, the variance in our estimated distances is only ${\sim}$4 kpc, or 5\% in distance. This provides us with an excellent empirical estimate of the distance precision of RR Lyrae candidates, before period-fitting (Sesar et al in prep). 
Note that many other satellites within ${\sim}$100 kpc also show clusters of RR Lyrae candidates (see Fig. \ref{fig:mollweide_rrlyrae}). 

\begin{figure*}[!ht]
 \begin{center}
\includegraphics{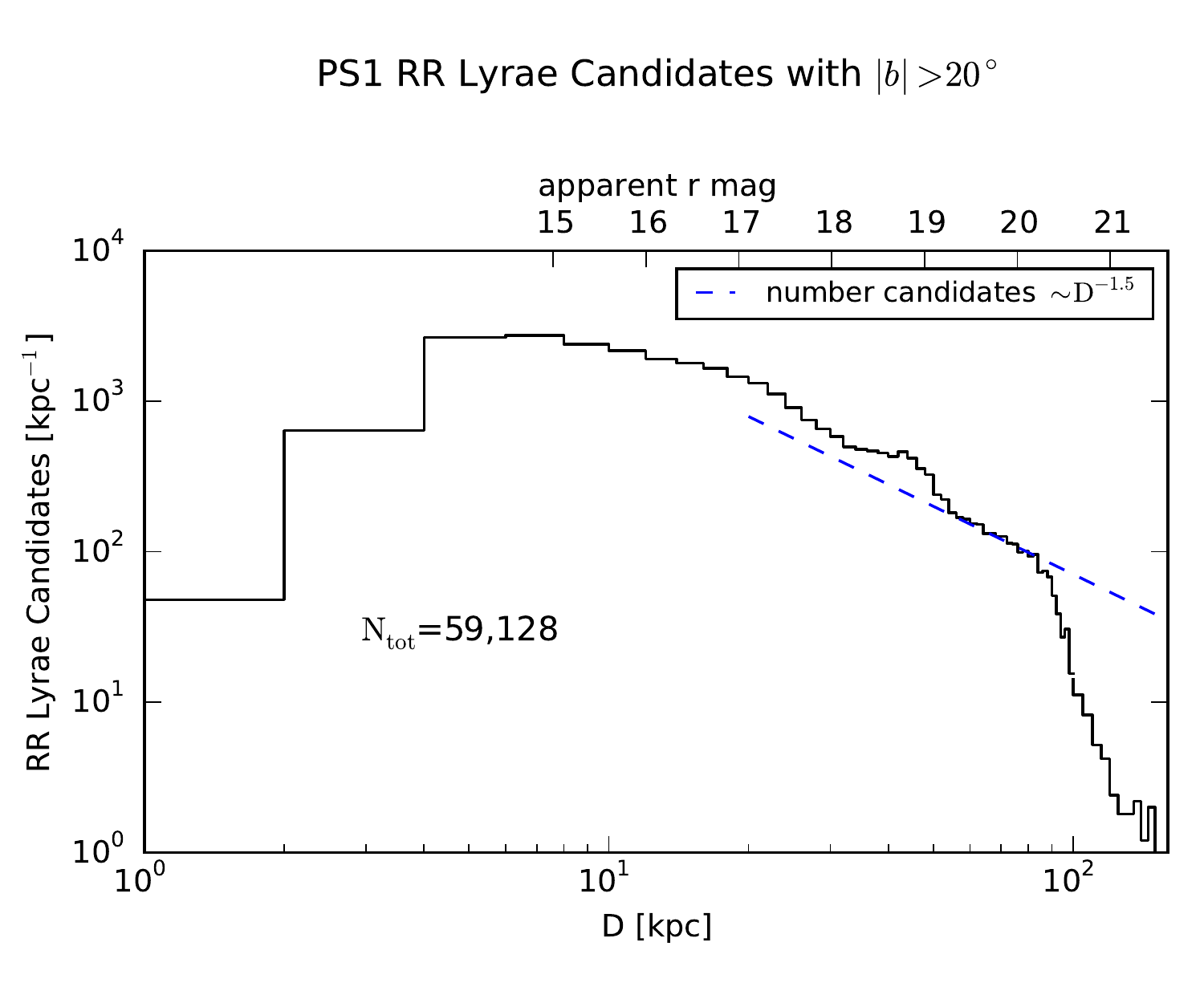}
                                \caption[]{\footnotesize{Distribution of the heliocentric distance estimates for 
                                halo RR Lyrae candidates ($p_{\mathrm{RRLyrae}} \geq 0.05$, $|b|>20\arcdeg$).
                                The corresponding apparent $r_{\rm P1}$ band magnitude, with no reddening assumed, is given.
                    Distance estimates are done from distance modulus of dereddened $r_{\rm P1}$ band mean magnitude. The 
figure shows the distances for the 59,128 out of 59,888 halo RR Lyrae candidates having $r_{\rm P1}$ band mean magnitude available. A number density profile ${\sim} \mathrm{D}^{-3.5}$ is overplotted.}} 
                \label{fig:hist_distance_halo_0p05}
                 \end{center}
\end{figure*}

In Fig. \ref{fig:hist_distance_halo_0p05} we show the heliocentric distance distribution of RR Lyrae candidates 
with $p_{\mathrm{RRLyrae}} \geq 0.05$ (purity=71\%, completeness=97\%) at Galactic latitudes $|b| \ge 20\arcdeg$. Half of them are within 20 kpc. The most distant candidates with $p_{\mathrm{RRLyrae}} \geq 0.05$ are ${\sim}$150 kpc away. 
A profile ${\sim} \mathrm{D}^{-1.5}$ related to a galactocentric halo density profile $\rho {\sim} \mathrm{D}^{-3.5}$ is overplotted for illustrative purposes. Such a halo profile is in the ball-park of recent determinations \citep{Xue2015, Deason2011, Sesar2010}. Comparing this profile to the distance distribution of our RR Lyrae candidates, we find this fits well up to ${\sim}80$ kpc. However, a rigorous derivation of the RR Lyrae density profile is beyond the scope of this paper.

\subsubsection{Comparison to the Catalina Survey}

Of course, PS1 is not the first large-area RR Lyrae survey at high Galactic latitudes; so in selected areas, we can also compare with previous surveys, e.g. SDSS \citep{York2000}, Catalina \citep{Drake2009}, QUEST \citep{Mateau2012}, and PTF \citep{Rau2009}.
Having used SDSS S82 in the training of the classifier, we focus here on the Catalina Sky Survey (CSS) \citep{Drake2009}, which 
has covered the region around the Galactic north pole down to $b=30\arcdeg$, but only for bright sources $\leq$ 19 mag.
CSS is a survey program for finding new
near-Earth objects, composed of the original Catalina Sky Survey (CSS),
the Siding Spring Survey (SSS) and the Mt. Lemmon Survey (MLS). Catalina photometry covers objects in the range -75$\arcdeg<\delta< 70\arcdeg$ and $|b| \gtrsim 15\arcdeg$. 
In addition to asteroid search, the complete Catalina data is analyzed for transient sources by the Catalina Real-time Transient Survey (CRTS), resulting in catalogs of RR Lyrae \citep{Drake2009, Drake2013a, Drake2013b}.
We use this to verify and check our RR Lyrae candidate identification, by cross-matching in this region with respect to CSS and SSS. We focus on the magnitude range in common between both surveys ${\sim} 15 - 18.5$ mag.

\begin{figure*}[H]            
\begin{center}    
\subfigure[ ]{\includegraphics{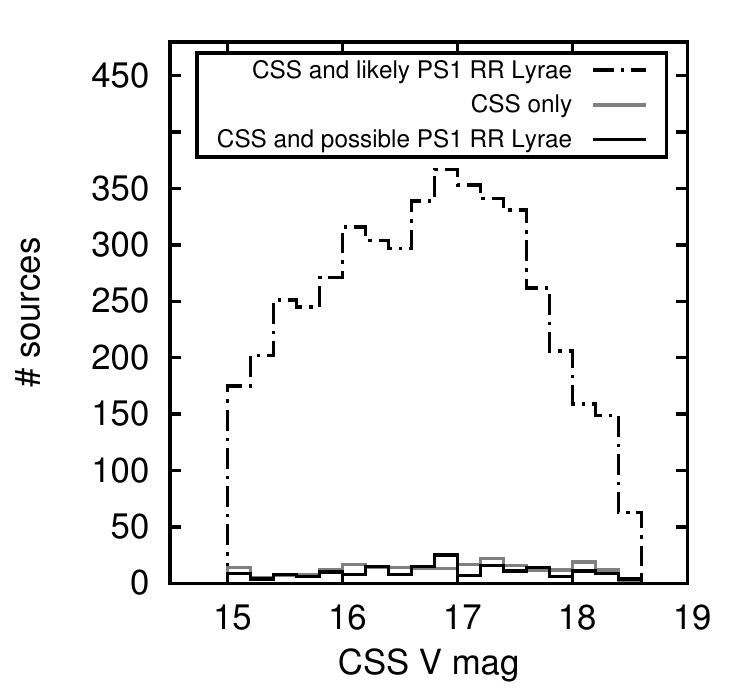}} 
\subfigure{\includegraphics{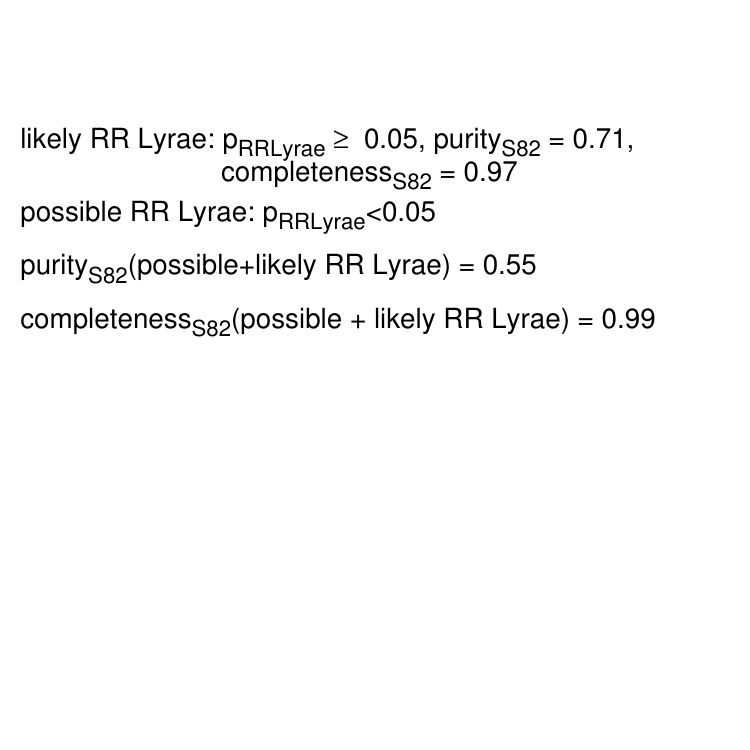}}
\setcounter{subfigure}{1}
\subfigure[ ]{\includegraphics{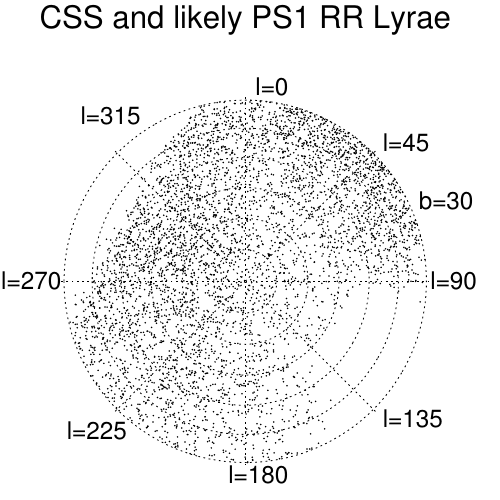}} 
\subfigure[ ]{\includegraphics{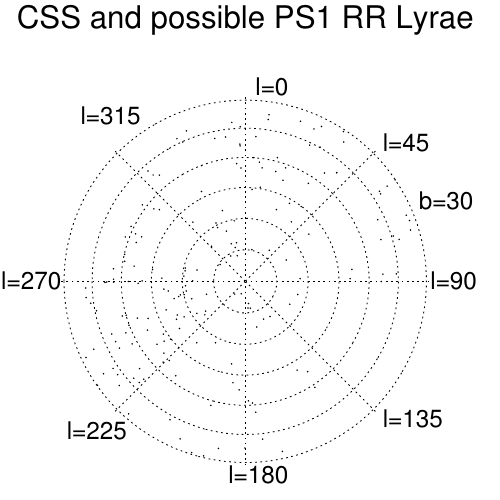}} 
\subfigure[ ]{\includegraphics{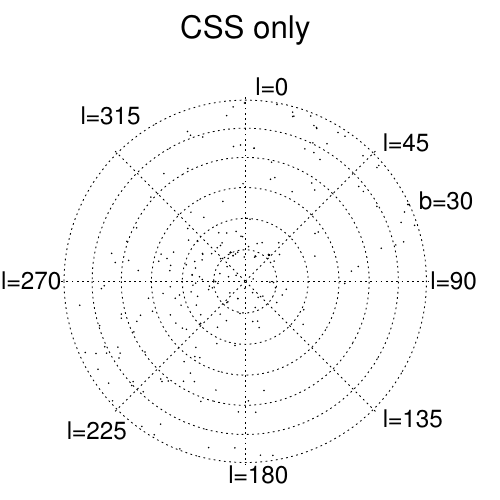}} 
\caption{\footnotesize{Distribution of CSS sources with $b>30\arcdeg$ cross-matched to PS1 sources within 5 arcsec. (b) The 4082 sources appearing in CSS and our classification as likely RR Lyrae having $p_{\mathrm{RRLyrae}} \geq 0.05$, (c) the 233 sources appearing in CSS and our classification as possible RR Lyrae having $p_{\mathrm{RRLyrae}} < 0.05$, (d) the 130 CSS sources that never enter our analysis. Panel (a) shows a histogram of the distribution of sources CSS V magnitude for the subsets from panels (b) to (d).}}
\label{fig:pRRLyrae_catalina_patch1}
 \end{center}
\end{figure*}

In Fig. \ref{fig:pRRLyrae_catalina_patch1}, we compare our RR Lyrae candidate list with the RR Lyrae identified by CSS. For this, we cross-match sources from our RR Lyrae candidate list to those of CSS with a matching distance of 5 arcsec and keep only the source with the closest match, following position errors reported in \cite{Casetti-Dinescu2015} which we also found for CSS. Additionally, we cut to a magnitude range covered by both CSS and our analysis, $15<V<18.5$.
In this magnitude range, the total number of CSS RR Lyrae with $b>30\arcdeg$ is 5108. For 4879 of them, we find a PS1 source within 5 arcsec. Out of these, 4686 have $p_{\mathrm{RRLyrae}} \geq 0.05$, 193 have $p_{\mathrm{RRLyrae}}<0.05$, and 229 never enter our analysis, as the PS1 source doesn't fulfill the selection criteria of Sec. \ref{sec:objectselectionandoutliercleaning}.

With respect to CSS, we get a completeness of 92\% (i.e. we find 92\% of their RR Lyrae),
and a cross-identified fraction of 30\% (i.e. they find 30\% of our sample), if we adopt the above magnitude cuts and $p_{\mathrm{RRLyrae}}$ threshold. The completeness of 92\% can be explained by the 10\% selection loss (Section \ref{sec:objectselectionandoutliercleaning}).

When comparing to the SSS RR Lyrae, we also chose a matching tolerance
of 5 arcsec and kept only the nearest match. Restricting to $15<V<18.5$, there are 3148 RR Lyrae in the region covered both by PS1 and SSS with $-30\arcdeg<\delta<-15\arcdeg$. Out of these, 2785 have $p_{\mathrm{RRLyrae}} \geq 0.05$, 233 have $p_{\mathrm{RRLyrae}} < 0.05$, and 130 never enter our analysis. To assess the cross-identified fraction, we have to consider $|b|>15\arcdeg$, as SSS roughly misses $|b|<15\arcdeg$. The number of PS1 RR Lyrae candidates in the
overlapping region and magnitude range and $p_{\mathrm{RRLyrae}} \geq 0.05$ not cross-matched to SSS is 11,336. The number of SSS RR Lyrae within these boundaries is 2725. 

In total, this results into a completeness with regard to SSS of 88\%, and a cross-identified fraction of 20\%.

We find 2 -- 3 times more RR Lyrae candidates with $p_{\mathrm{RRLyrae}} \geq 0.05$ than the pure samples of CSS and SSS RR Lyrae. This is in line with our assessment of purity at such lenient candidate criteria $p_{\mathrm{RRLyrae}} \geq 0.05$. 
Taken together, CSS and SSS's claim of 70\% completeness \citep{Torrealba2009}, our 10\% selection loss as of Sec. \ref{sec:objectselectionandoutliercleaning} and our purity of ${\sim}71\%$
at $p_{\mathrm{RRLyrae}} \geq 0.05$, we expect about 44\% of our candidates to be cross-identified in CSS or SSS; this is close to the actual fraction of 30\% for CSS within 5 arcsec. In the SSS, we obtain a lower cross-identified fraction of 20\%; this suggests that the completeness of the SSS is in fact lower than that of the CSS.

\subsection{The Catalog of Variable Sources in PS1 3$\pi$}
\label{sec:Catalog}

We have processed $3.88 \times 10^8$ PS1 3$\pi$ sources that fulfill the cuts described in Sec. \ref{sec:PS13pidata}.
From these, we provide a catalog of all likely variable point sources in PS1 and of all likely QSOs,
a total of 25,790,103 sources. We include all sources fulfilling the criterion of $\log \hat{\chi}^2>0.5$ (see Fig. \ref{fig:chihatdistribution}) or $W12>0.5$. The latter criterion is intended to ensure that we provide variability statistics for almost all QSOs.

The Catalog of Variable Sources is available in its entirety in machine-readable format in the online journal. A table structure is shown here for guidance regarding its form and content.

\def\arraystretch{1.5}

\begin{deluxetable*}{r l l}
\tabletypesize{\scriptsize}
\tablecaption{The Catalog of Variable Sources in PS1 3$\pi$}
\label{tab:catalog}
\tablewidth{0pt}
\tablehead{
\colhead{Column} & \colhead{FITS Format Code} & \colhead{Description}  }
\startdata
1 & E &   right ascension in degrees\\
2 & E &   declination in degrees \\
3 & E &    scalar variability quantity $\hat{\chi}^2$, Equ. \eqref{eqn:q}\\
4 & E &   best fit structure function parameter $\omega_r$ ($r$ band variability amplitude) on log-spaced grid\\
5 & E &   best fit structure function parameter $\tau$ (time scale) on log-spaced grid\\
6 & E &   error-weighted mean $g_{\rm P1}$ band magnitude $\langle g_{\rm P1} \rangle $ \\ 
7 & E &  error-weighted mean $r_{\rm P1}$ band magnitude $\langle r_{\rm P1} \rangle $ \\ 
8 & E &  error-weighted mean $i_{\rm P1}$ band magnitude $\langle i_{\rm P1} \rangle $ \\ 
9 & E &  error-weighted mean $z_{\rm P1}$ band magnitude $\langle z_{\rm P1} \rangle $ \\ 
10 & E &  error-weighted mean $y_{\rm P1}$ band magnitude $\langle y_{\rm P1} \rangle $ \\ 
11 & E &  W1-W2 color from WISE\\
12 & E & $p_{\mathrm{QSO}}$ \\    
13 & E & $p_{\mathrm{RRLyrae}}$
\enddata
\tablecomments{Structure of the Catalog of Variable sources in PS1 3$\pi$. The Catalog of Variable sources is available in its entirety in machine-readable format in the online journal. A table structure is shown here for guidance regarding its form and content.
}
\end{deluxetable*}

\section{Discussion and Conclusion}
\label{sec:DiscussionandConclusion}

We have set out to identify, characterize and classify variable (point) sources in the PS1 survey, the most extensive, deep, multi-band, wide-area, multi-epoch imaging survey to date.
Because photometry in different bands of PS1 are not observed simultaneously (as they were e.g. in SDSS), we had to develop and implement new methodology for multi-band fitting
of structure functions, used to characterize non-simultaneous multi-band lightcurves. This allowed us to assign to each of almost half a billion point sources in PS1 a basic, $\chi^2$-based 
variability indicator, a variability amplitude (in the $r_{\rm P1}$-band) $\omega_r$, and a variability time-scale $\tau$. 

We then focused on identifying two classes of variable sources among these objects, QSOs and RR Lyrae stars. Because it aids enormously in the identification of QSOs, we also 
matched all sources to band $W1$ and $W2$ photometry from the WISE space mission. To classify objects on the basis of this mean photometry and lightcurves, we exploited the fact that 
SDSS Stripe 82 is covered by PS1, and provides a full inventory of QSOs and RR Lyrae in that area. We take S82 classification (QSO, RR Lyrae and ``other'') as ground truth, to train a Random Forest Classifier to 
classify all sources in PS1 that are brighter than 21.5 mag in either $g_{\rm P1}$, $r_{\rm P1}$, or $i_{\rm P1}$. 
  
We have not only carried out classification with the full available parameter set using variability parameters and colors from PS1 together with WISE colors, but also with more restricted pieces of information, using only color-related and only variability-related parameters.
For RR Lyrae, the variability information is absolutely indispensible to define a sample with an interesting combination of purity and completeness. 

For QSOs, (time-averaged) PS1 color together with WISE already do a good job in selecting QSOs, so PS1 variability provides a significant, but not decisive improvement of purity and completeness. On the other hand, this means, that the variability information together with optical color can help for QSO identification when no other information is available.

As the treatment of reddening is limited right now, care must be taken applying any values of purity and completeness to regions of high reddenings.

One important limitation of our classification is that it relies on SDSS Stripe 82; while this area covers a wide range in Galactic latitude, $20\arcdeg < b < 70\arcdeg$, we have no training in the galactic plane. While the number of very likely candidates, $p_{\mathrm{QSO/RRLyrae}}>0.2$ drops near the galactic plane, the number of plausible candidates $p_{\mathrm{QSO/RRLyrae}}>0.05$ does not. This implies, unsurprisingly, that we are likely to have considerably higher contaminations, at least in the
$p_{\mathrm{QSO/RRLyrae}}>0.05$-sample than our tests in S82 would imply. The purity of low-latitude samples must be settled with follow-up observations and analysis.  
However, at high galactic latitudes, 
PS1 appears to remain quite complete in its selection to nearly
$r_{\rm P1}{\sim} 21$, which enables candidate selection to nearly ${\sim} 140$ kpc.

Across the entire $3\pi$, we identified 247,281 RR Lyrae candidates in PS1 with
$p_{\mathrm{RRLyrae}} \geq 0.05$.
Based on the training in S82, we expect a purity (based on S82)
of 71\%, and completeness of 98\%  among cross-matched sources;
10\% of the sources will be missing because of the selection loss (see Sec. \ref{sec:objectselectionandoutliercleaning}). As mentioned above, these numbers on purity and completeness only apply away from the Galactic plane, and the bulge.
Increasing the threshold to the more stringent criteria of $p_{\mathrm{RRLyrae}} \geq
0.2$, reduces the sample to 153,151 sources.

The S82 training would make us believe that this should boost the purity
to 75\%  with only a slightly lower completeness of 92\%.
The fact that nearly 100,000 candidates fall out of the sample between the
two cuts of ($p_{\mathrm{RRLyrae}} \geq 0.05$ and $p_{\mathrm{RRLyrae}} \geq 0.2$)
shows that the purity in the $p_{\mathrm{RRLyrae}} \geq 0.05$ sample must be overestimated.
This is most likely because there is not only dust, but also higher, and unmodelled, contamination in the Galactic plane.

With this caveat on the low-latitude sample purity, the spatial distribution of RR Lyrae
candidates is as follows: 
Within $|b| < 20 \arcdeg$, i.e. near the disk, we find 187,393 possible RR Lyrae candidates with $p_{\mathrm{RRLyrae}} \geq 0.05$ and 
110,477 RR Lyrae candidates with $p_{\mathrm{RRLyrae}} \geq 0.2$. Of them, 19,958 with $p_{\mathrm{RRLyrae}} \geq 0.05$ and 12,967 with $p_{\mathrm{RRLyrae}} \geq 0.2$ may be 
in the bulge as being in a radius of 20$\arcdeg$ around the Galactic center.  Here we refer to the selection cuts in the parameter $p_{\mathrm{RRLyrae}}$, because the mapping to purity and completeness in S82 may not apply at such low latitudes. In the Galactic halo, at Galactic latitudes of $|b| >  20\arcdeg$ we have 59,888 candidates with $p_{\mathrm{RRLyrae}} \geq 0.05$, some extending to distances as large as ${\sim}$140 kpc.

This is the most extensive and faintest RR Lyrae candidate sample to date, extending to considerably fainter magnitudes than e.g. the CRTS sample of RR Lyrae stars.
Using the RR Lyrae in Draco, we show that distances derived from $\langle r_{\rm P1} \rangle$ and $M_r=0.6$ we get distance precisions of 6\% at a distance of ${\sim}$80 kpc. 
A projection of our candidate sample into the orbital plane of the Sagittarius stream reveals the stream morphology clearly.
This shows that this sample will be excellent for mapping stellar (sub-)structure in the Galactic halo.

We have selected 399,132 likely QSO candidates over the total PS1 3$\pi$ area at a level of purity of 82\%, completeness of 75\%, and 1,596,319 possible candidates at a level of purity of 72\%, completeness of 98\%. The selection of candidates is homogeneous to a high degree away from the Galactic plane. 
At $|b| > 20\arcdeg$, we find 784,233 candidates with $p_{\mathrm{QSO}} \geq 0.2$ and 356,732 candidates with $p_{\mathrm{QSO}} \geq 0.6$.
The selection of candidates is homogeneous to a high degree away from the Galactic plane. Around the plane, the number density of QSO candidates with high $p_{\mathrm{QSO}}$ decreases because of dust.

Over all, this work has resulted in estimation of variability parameters and mean magnitudes for more than $3.88 \times 10^8$ sources, and a catalog of variable sources of almost $2.58 \times 10^ 7$ objects, being available as a 3$\pi$ value-added catalog. These parameters of course allow the source classification based on different training sets than the one presented here.

These results of PS1 3$\pi$
variability studies in the MW context offer the possibility for all-sky detection of variable sources and will enable us to use RR Lyrae to precise distance estimates for finding streams and satellites. QSO candidates will be used
as a reference frame for Milky Way astrometry, to get absolute proper motions
and study Milky Way disk kinematics. 

Candidates of periodic variables can be processed further to increase their purity. As approaches for period finding and fitting are very computational expensive, it needs to be applied to pre-selected candidates 
\citep[see][Sesar et al. in prep.]{VanderPlas2015}.

Several approaches for detecting period lightcurve signals exist for well-sampled single-band data \citep[e.g.][]{Sesar2010,Graham2013}, 
but must be adopted for the randomly sampled multiband lightcurves as present from PS1 and LSST.
Promising approaches for detecting periodicity in sparsely sampled multi-band time domain data are the multiband periodogram \citep{VanderPlas2015} as well as lightcurve template fitting (Sesar et al. in prep.).

Looking forward to catalogs of variable stars from Pan-STARRS, LSST and other multi-band all-sky time-domain surveys, our approach meets the constraints of being able to deal with noisy observational through different bands, accompanied by data from other surveys, and is fast enough to provide a sample pure and complete enough for further lightcurve analysis.

\section{Acknowledgments}

The research leading to these results has received
funding from the European Research Council
under the European Union's Seventh Framework
Programme (FP 7) ERC Grant Agreement n.
[321035].

HWR, ES and EKG acknowledge funding by the Sonderforschungsbereich SFB 881 The Milky Way System (subproject A3) of the German Research Foundation (DFG). N.F.M. gratefully acknowledges the CNRS for support through PICS project PICS06183.

The Pan-STARRS1 Surveys (PS1) have been made possible through contributions of the Institute for Astronomy, the University of Hawaii, the Pan-STARRS Project Office, the Max-Planck Society and its participating institutes, the Max Planck Institute for Astronomy, Heidelberg and the Max Planck Institute for Extraterrestrial Physics, Garching, The Johns Hopkins University, Durham University, the University of Edinburgh, Queen's University Belfast, the Harvard-Smithsonian Center for Astrophysics, the Las Cumbres Observatory Global Telescope Network Incorporated, the National Central University of Taiwan, the Space Telescope Science Institute, the National Aeronautics and Space Administration under Grant No. NNX08AR22G issued through the Planetary Science Division of the NASA Science Mission Directorate, the National Science Foundation under Grant No. AST-1238877, the University of Maryland, and Eotvos Lorand University (ELTE) and the Los Alamos National Laboratory. 

Funding for SDSS-III has been provided by the Alfred P. Sloan Foundation, the Participating Institutions, the National Science Foundation, and the U.S. Department of Energy Office of Science. The SDSS-III web site is http://www.sdss3.org/.

SDSS-III is managed by the Astrophysical Research Consortium for the Participating Institutions of the SDSS-III Collaboration including the University of Arizona, the Brazilian Participation Group, Brookhaven National Laboratory, Carnegie Mellon University, University of Florida, the French Participation Group, the German Participation Group, Harvard University, the Instituto de Astrofisica de Canarias, the Michigan State/Notre Dame/JINA Participation Group, Johns Hopkins University, Lawrence Berkeley National Laboratory, Max Planck Institute for Astrophysics, Max Planck Institute for Extraterrestrial Physics, New Mexico State University, New York University, Ohio State University, Pennsylvania State University, University of Portsmouth, Princeton University, the Spanish Participation Group, University of Tokyo, University of Utah, Vanderbilt University, University of Virginia, University of Washington, and Yale University. 

The CSS survey is funded by the National Aeronautics and Space Administration under Grant No. NNG05GF22G issued through the Science Mission Directorate Near-Earth Objects Observations Program. The CRTS survey is supported by the U.S. National Science Foundation under grants AST-0909182 and AST-1313422. The services at IUCAA are supported by the University Grants Commission and the Ministry of Information Technology, Govt. of India under the Virtual Observatory - India project.

We acknowledge early contributions to the implementation of the structure function formalism of David Mykytyn (NYU) and Ekta Patel (Arizona).

The Figure \ref{fig:triangleplot} was created with {\tt corner.py} v1.0.0 by Foreman-Mackey, D., Price-Whelan, A., Ryan. G., et al.

\clearpage
\appendix

\section{FIGURES}
\label{sec:Appendix}

 \begin{figure*}[!ht]
              \includegraphics[angle=90,origin=c,height=0.8\textheight]{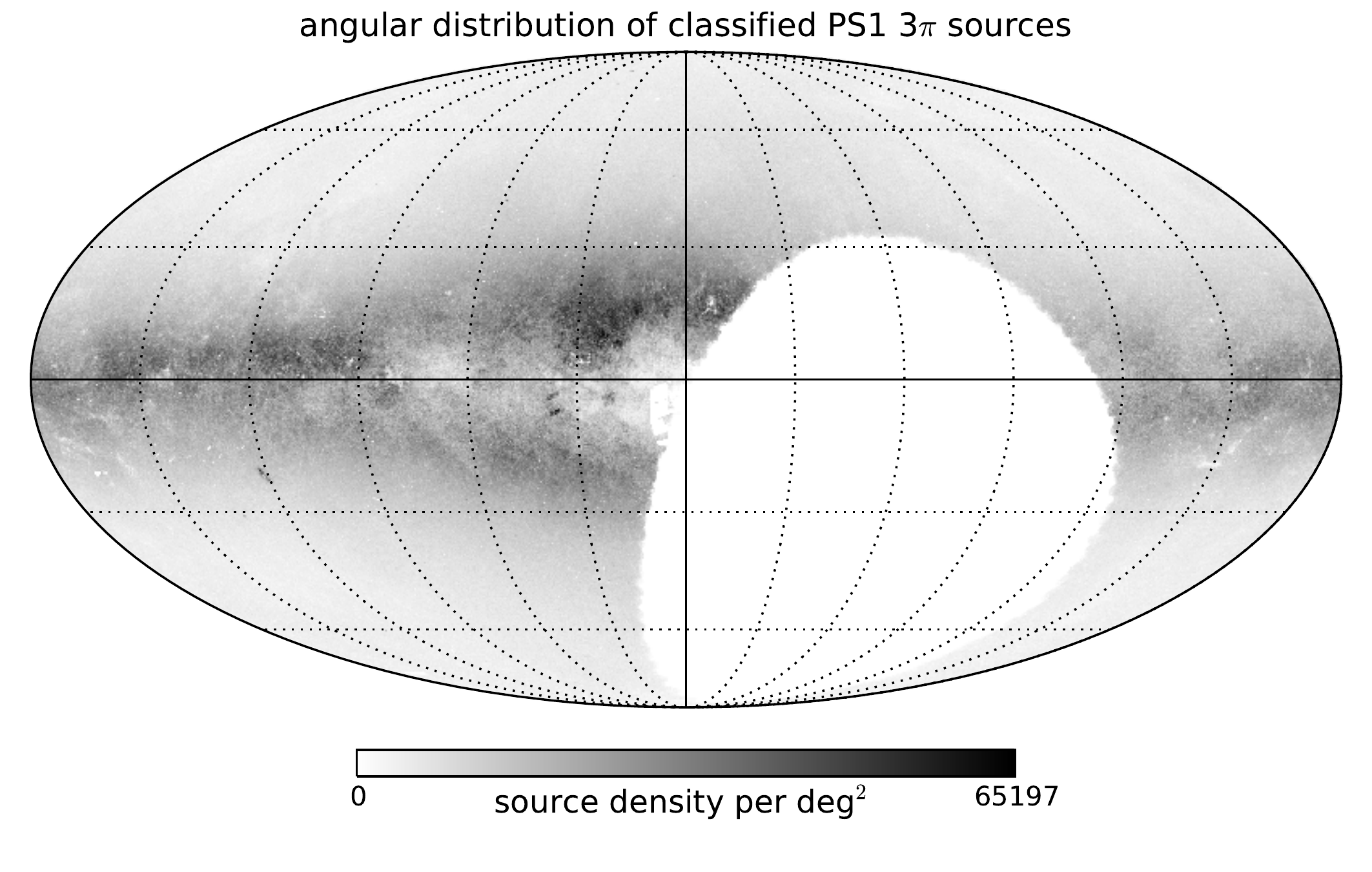} 
                \caption[shortcaption]{\footnotesize{Density of processed $3.88 \times 10^8$ PS1 3$\pi$ sources as Mollweide projection in Galactic coordinates using the healpy (https://healpy.readthedocs.org) pixelation.}}
                \label{fig:healpix_map}
\end{figure*}

 \begin{figure*}[!ht]
              \includegraphics[angle=90,origin=c,height=0.8\textheight]{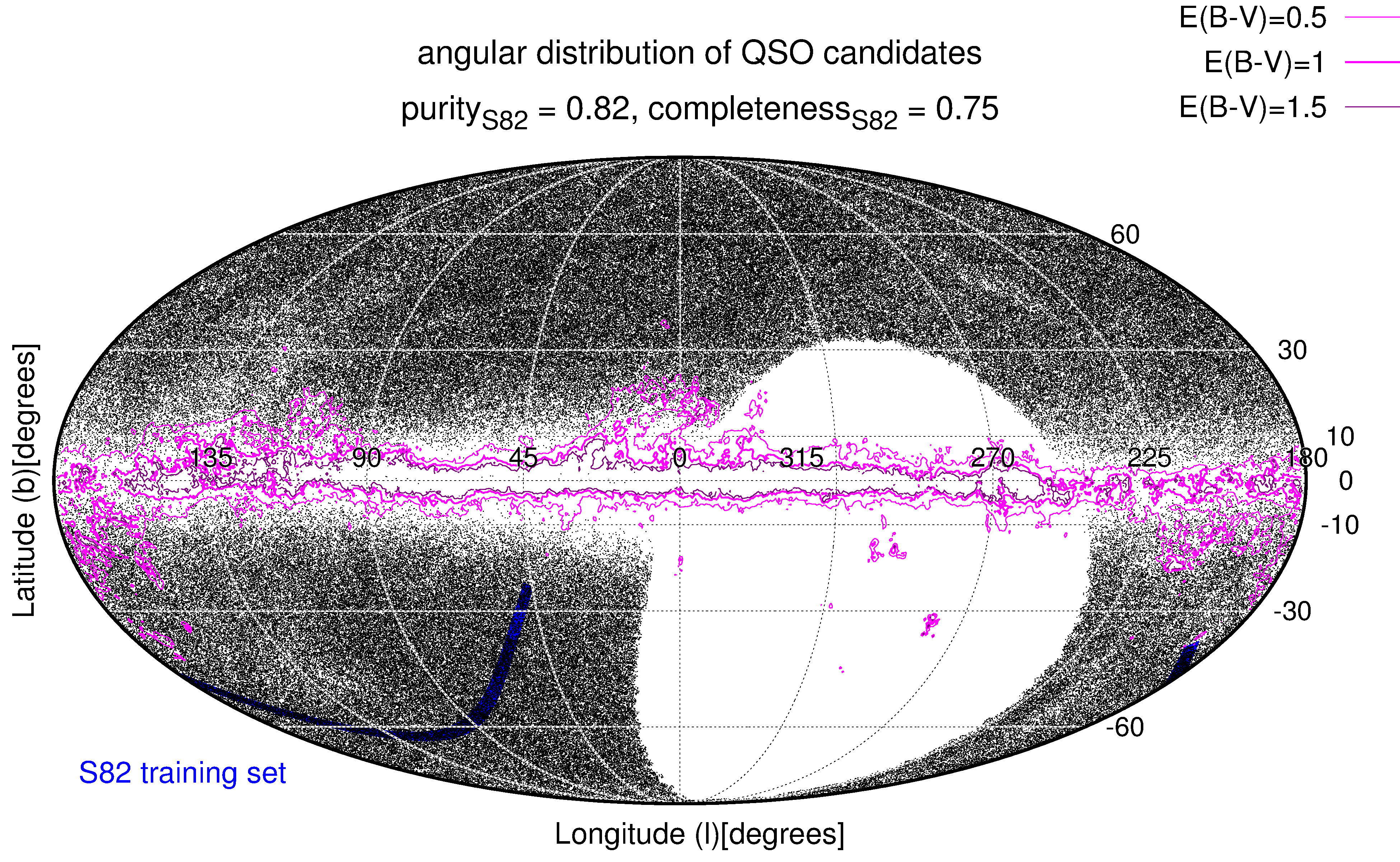}  
                \caption[distribution of QSO candidates ($0.6 \leq p_{\mathrm{QSO}} \leq 1$]{\footnotesize{Distribution of the 399,132 QSO candidates ($0.6 \leq p_{\mathrm{QSO}} \leq 1$, purity=82\%, completeness=75\%), shown in Mollweide projection in Galactic coordinates. A contour plot of the reddening-based $E(B-V)$ dust map \citep{Schlafly2014} is overlayed.}}
                \label{fig:mollweide_qso}
\end{figure*}

 \begin{figure*}[!ht]
              \includegraphics[angle=90,origin=c,height=0.8\textheight]{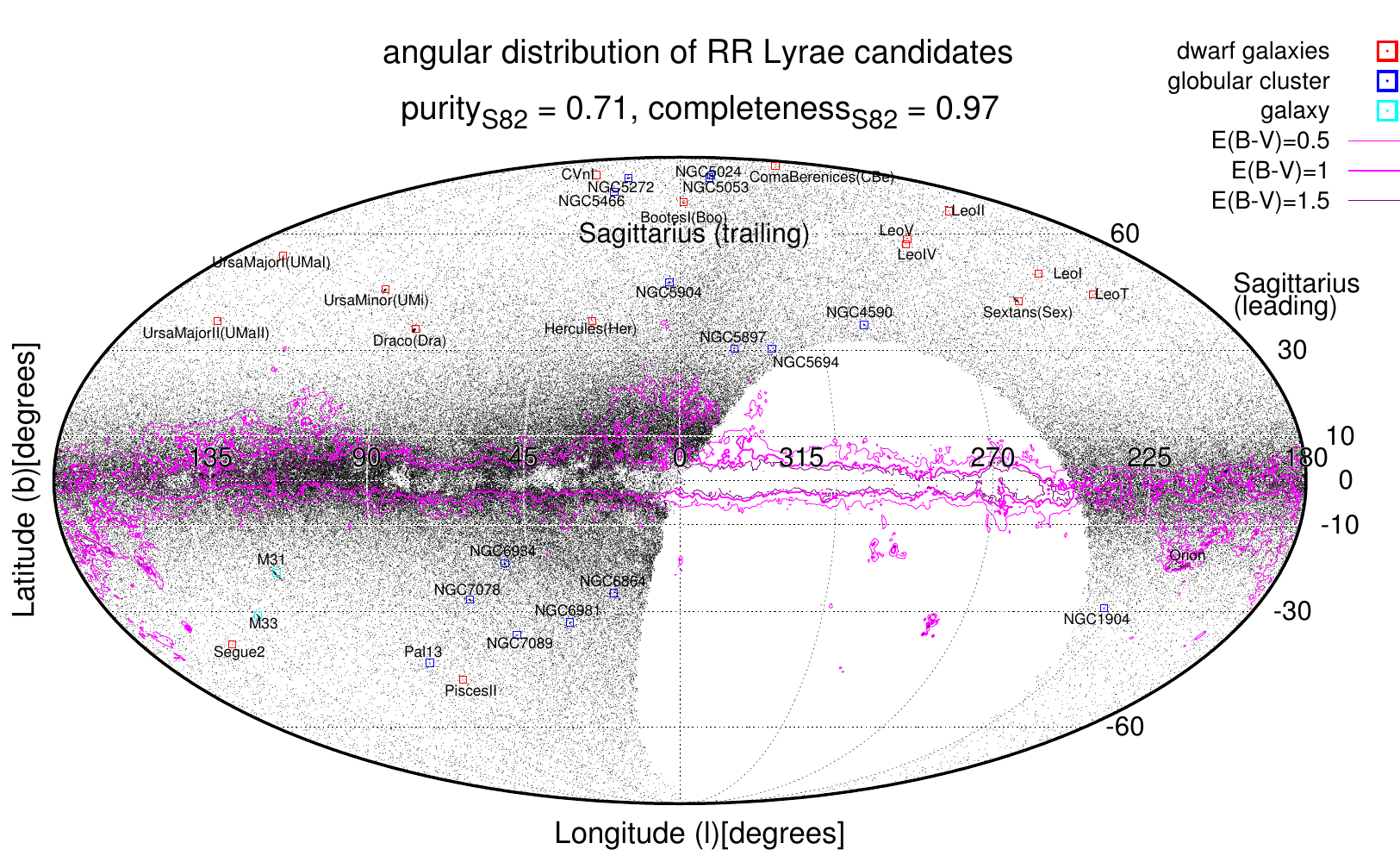}  
                                \caption[distribution of RR Lyrae candidates ($p_{\mathrm{RRLyrae}} > 0.05$]{\footnotesize{Distribution of the 247,281 RR Lyrae candidates ($p_{\mathrm{RRLyrae}} > 0.05$, purity=71\%, completeness=97\%), shown in Mollweide projection of Galactic coordinates. A contour plot of the reddening-based $E(B-V)$ dust map \citep{Schlafly2014} is overlayed, as well as identified known objects of the Milky Way spheroid substructure and its neighborhood.}} 
                \label{fig:mollweide_rrlyrae}
\end{figure*}

\end{document}